\documentclass[amsmath,amssymb,11pt]{article}
\usepackage{jheppub2}
\usepackage{tikz}
\usepackage{xcolor}
\pdfoutput=1

\usepackage{amsmath,amssymb,amsthm,mathrsfs,bbm,bm}
\usepackage{latexsym,amscd,amsbsy,amsfonts,dsfont}
\usepackage{multirow}
\usepackage{graphicx}
\usepackage{xcolor}
\usepackage{caption}
\usepackage{comment}

\newcommand{\beq}{\begin{equation}}
\newcommand{\eeq}{\end{equation}}
\newcommand{\beqa}{\begin{eqnarray}}
\newcommand{\eeqa}{\end{eqnarray}}
\newcommand{\bea}{\begin{eqnarray}}
\newcommand{\eea}{\end{eqnarray}}

\newcommand*{\affmark}[1][*]{\textsuperscript{#1}} 

\numberwithin{equation}{section}  
\allowdisplaybreaks

\title{Love at First Loop: Quantum Deformability of Near-Extremal Black Holes}

\author{Pablo A. Cano\affmark[1],}
\emailAdd{pablocano@um.es}
\author{Marina David\affmark[2,3,4],}
\emailAdd{marina.claudia.david@cern.ch}
\author{Roberto Emparan\affmark[5,6],}
\emailAdd{emparan@ub.edu}
\author{Stefano Trezzi\affmark[6]}
\emailAdd{strezzi@icc.ub.edu}

\affiliation{
\affmark[1]Departamento de Física, Universidad de Murcia, Campus de Espinardo, 30100 Murcia, Spain\\
\affmark[2]Theoretical Physics Department, CERN, 1211 Geneva 23, Switzerland\\
\affmark[3]Institute for Theoretical Physics, KU Leuven,
Celestijnenlaan 200D, B-3001 Leuven, Belgium\\
\affmark[4]Leuven Gravity Institute, KU Leuven, Celestijnenlaan 200D, B-3001 Leuven, Belgium\\
\affmark[5]Institució Catalana de Recerca i Estudis Avançats (ICREA),
 Passeig Lluis Companys, 23, 08010 Barcelona, Spain\\
\affmark[6]Departament de Física Quàntica i Astrofísica and
  Institut de Ciències del Cosmos (ICCUB), Martì i Franquès~1,
 Universitat de Barcelona, 08028 Barcelona, Spain\\
}

\abstract{
The static Love numbers of black holes vanish in classical, four-dimensional, asymptotically flat General Relativity. We study how this changes for the response of near-extremal Reissner--Nordstr\"om black holes to a minimally coupled massless scalar, at energies where the Schwarzian mode of the near-horizon throat fluctuates strongly. The black hole then responds as a quantum system with a spectrum of states. Its static response is a polarizability, given by virtual transitions between them. For $\ell=0$ the static Love number vanishes, by the shift symmetry of the scalar, but for multipoles $\ell\geq1$ the static Love numbers are nonzero: the finite rotational energy of the spin-$\ell$ intermediate states spoils the cancellation responsible for the classical result. Since the throat no longer has a classical geometry from which to read off a tidal response, we define it through the retarded Green's function, and compute its full frequency dependence at leading order in $r_+\omega$. Its conservative part requires renormalization: 
the near-horizon theory fixes the coefficient of its logarithm, and with it a dependence on the energy that no local term can remove,
while its finite local terms are matching data of the four-dimensional theory. The dissipative part is finite and scheme independent. For $\ell\geq1$, it vanishes at low frequencies when the energy is below the rotational gap, while at the gap threshold it becomes non-analytic in frequency and produces a late-time $t^{-3/2}$ tail.
}

\begin{document}

\preprint{CERN-TH-2026-249}

\maketitle

\section{Introduction and Summary}
\label{sec:intro}

How black holes respond to external perturbations is a fundamental question that probes their nature beyond the classical no-hair theorem. Love numbers---the dimensionless coefficients encoding the distortion of a body in an external tidal field---provide a sharp characterization of this response, and their measurement offers direct information about the structure and dynamics of the object being probed \cite{Binnington:2009bb,Damour:2009vw,Kol:2011vg,Hui:2020xxx}.\footnote{See \cite{Rodriguez:2026iot,Chakraborty:2026qru} for recent reviews.}

For black holes in four-dimensional, asymptotically flat General Relativity, the static Love numbers vanish at the classical level \cite{Binnington:2009bb,Damour:2009vw,Kol:2011vg,Hui:2020xxx,LeTiec:2020bos,LeTiec:2020spy,Chia:2020yla,Ivanov:2022qqt}. This remarkable property has been related to hidden or emergent symmetries of the perturbation equations, often involving $SL(2,\mathbb{R})$, with different approaches identifying different realizations of it \cite{Charalambous:2021mea,Charalambous:2022rre,Hui:2021vcv,Kehagias:2024rtz,Gounis:2024hcm,Lupsasca:2025pnt,Parra-Martinez:2025bcu}.

In this article we study whether the situation changes when quantum effects are taken into account, in particular for black holes near extremality, where these effects can be large yet controllable. Recent work has shown that quantum loops make the scalar and electromagnetic Love numbers of Schwarzschild black holes run logarithmically, so that they cannot vanish, at order $\ell_{\rm P}^2/r_h^2$ \cite{Elkhidir:2026lgp} (see also \cite{Kim:2020dif}).\footnote{
Nonzero Love numbers also arise from changing the asymptotic boundary conditions \cite{Emparan:2017qxd,Franzin:2024cah}, from higher-derivative corrections \cite{Cardoso:2018ptl,DeLuca:2022tkm,Barbosa:2025uau,Cano:2025zyk,Barbosa:2026qcv,Noumi:2026shc}, and for string stars near the correspondence with black holes \cite{Emparan:2026wlm}.
} Our focus here is instead on the universal strong quantum effects close to extremality. Near-extremal Reissner--Nordstr\"om black holes develop a long near-$\text{AdS}_2$ region whose low-energy gravitational dynamics is governed by the Schwarzian action. At low enough energies its quantum fluctuations become parametrically large, while remaining under control within the Schwarzian framework \cite{Iliesiu:2020qvm,Mertens:2022irh}.

The black hole is then no longer described as a fixed background, but
as a quantum system with a spectrum of states. Its static response is
a polarizability, given by a sum over virtual transitions between these states.
A tidal field with $\ell\geq1$ connects an initially spinless black hole to intermediate states of spin $\ell$. The rotational energy of these states, which is negligible in the classical regime, upsets the balance between contributions from higher and lower energies, leaving a nonzero static response. The effect is absent for $\ell=0$, where no angular momentum is transferred.

A number of recent studies have examined how asymptotic observers can probe these fluctuations through Hawking emission spectra, scattering experiments that measure the absorption cross-section, quasinormal modes and late-time decay, or black hole-induced decoherence of quantum superpositions \cite{Bai:2023hpd,Brown:2024ajk,Maulik:2025hax,Emparan:2025sao,Biggs:2025nzs,Lin:2025wof,Li:2025vcm,Emparan:2025qqf,Bhattacharjee:2025wfv,Biggs:2026zlp,Emparan:2026ta}. All these observables are ultimately sensitive to near-extremal two-point functions, which are significantly modified by the Schwarzian dynamics relative to their classical counterparts. Here we study the response of the near-extremal throat to an external, minimally coupled, neutral massless scalar. Our central object is the full Schwarzian-dressed retarded response function. Its conservative and dissipative parts contain, respectively, the static tidal response and the net absorptive response. The latter includes stimulated emission, which is forbidden when the final black hole energy would lie below the minimum allowed in the relevant spin sector.

\paragraph{Two formulations of tidal response.}
Already at the classical level there are two ways of computing this response. In the conventional calculation, one solves the perturbation equation with the appropriate boundary condition at the horizon and reads the applied field and the induced response from the coefficients of the two asymptotic radial branches \cite{Binnington:2009bb,Damour:2009vw,Kol:2011vg}. Their ratio is the response function $\hat\chi_\ell$.

In the other formulation, the applied field is treated as an external source and the response is defined through a renormalized generating functional. Its first variation gives the induced expectation value, and the linear dependence of that expectation value on the source is the retarded two-point function that governs the linear response. The divergences are removed by local boundary counterterms, with the usual freedom to add finite local terms.

The two formulations are related as in holographic renormalization. Reading a Love number from a radial coefficient is analogous to reading a one-point function from a coefficient of the Fefferman--Graham expansion, where the subtraction of divergences is built into the prescription \cite{Balasubramanian:1998de,Klebanov:1999tb,deHaro:2000vlm}, while the generating functional is the analogue of the covariant formulation of ref.~\cite{Balasubramanian:1999re}. As in AdS/CFT, the two agree up to local terms. The relation between Love numbers and holographic two-point functions was made explicit in ref.~\cite{Emparan:2017qxd}.

The distinction matters once the near-horizon geometry itself fluctuates quantum mechanically. There is then no classical geometry, nor a definite radial profile, from which source and response coefficients can be read, whereas the generating functional remains well defined after integrating over the Schwarzian mode. We therefore use the renormalized retarded correlator as the definition of the quantum response.

\paragraph{Setup: the response of the near-extremal throat.}
We work in the microcanonical ensemble at fixed energy above extremality $E=M-Q$, with an initially spinless black hole. The scale
\begin{equation}
    E_b=\frac{\pi}{r_+ S_0}\,,
\end{equation}
with $S_0$ the (naive) extremal entropy, controls the onset of strong Schwarzian effects when $E$ becomes of order $E_b$ or smaller. We compute the retarded Green's function at the mouth of the throat and normalize the response according to
\begin{equation}
    \hat\chi_\ell(\omega)
    =\mathcal{N}_\ell\,\widehat{G}^{R,\mathrm{ren}}_\ell(\omega),
\end{equation}
where $\mathcal{N}_\ell$ is fixed by matching to the classical response. We work at leading order in $r_+\omega\ll 1$, keeping the exact dependence on $\omega/E_b$ and $E/E_b$. At the order considered, the response at the mouth of the throat coincides with that in the intermediate region of the asymptotically flat geometry; corrections are suppressed by $r_+\omega$ and will not be included.

A scalar perturbation of angular momentum $\ell$ connects the initially spinless black hole to the spin-$\ell$ sector, whose spectrum starts at
\begin{equation}
    E_{0,\ell}=\frac{\ell(\ell+1)}{2}\,E_b\,.
\end{equation}
This is the rotational energy referred to above.

\paragraph{Conservative response: renormalization and running.}
The conservative part of the quantum response has ultraviolet divergences and requires renormalization. Divergences also arise when the classical problem is formulated through the boundary action, but there, the $\text{AdS}_2$ scaling symmetry leaves no finite local term that can shift the response. The Schwarzian scale $E_b$ changes this. For $\ell\geq1$ the response has a logarithmic divergence, which is part of the static Love number.
With the finite local terms held fixed, the static Love number then runs with the renormalization scale $\mu$,
\begin{equation}
\label{eq:lambda_ell_RG_intro}
\lambda_\ell(\mu)
= \lambda_\ell(\mu_0)
+\text{B}_\ell \log\frac{\mu}{\mu_0}.
\end{equation}
The value $\lambda_\ell(\mu_0)$ at a reference scale $\mu_0$ depends on the scheme, but the beta function $\text{B}_\ell$ does not. 
For $\ell=1$ it is
\begin{equation}
\label{eq:B1_intro}
\text{B}_1
= - \frac{(r_+ E_b)^3}{12 \pi}
\left(\frac{8 E}{E_b} + 1 \right),
\end{equation}
and the results for higher multipoles are given in \eqref{eq:B_ell_1} and \eqref{eq:B1}.

Since $\text{B}_\ell$ is a polynomial in the energy, a change of $\mu$ can be compensated by a change of the finite local terms, and the physical response of a given black hole does not depend on $\mu$. The running acquires physical content only through the scale that $\mu$ is compared with, which in the static response is the larger of the rotational energy $E_{0,\ell}$ and $\sqrt{2E_bE}$. At high energies, where $\sqrt{2E_bE}\gg E_{0,\ell}$, the Love number depends on $\log E$, with coefficient $-\text{B}_\ell/2$; at lower energies, instead, $\mu$ is compared with $E_{0,\ell}$, so the logarithm does not depend on $E$. In both cases the full dependence on the energy is not polynomial, so the local terms cannot absorb it, and the Love number cannot vanish at all energies.

With the conventional normalization in terms of the horizon radius, these Love numbers are very small. Since $r_+E_b=\pi/S_0$, for $E\sim E_b$ they are of order $S_0^{-(2\ell+1)}$, well below the generic loop corrections of ref.~\cite{Elkhidir:2026lgp}, which are of order $S_0^{-1}$.\footnote{However, the results of \cite{Elkhidir:2026lgp} apply to Schwarzschild black holes, so they cannot be directly compared to the near-extremal Reissner-Nordstr\"om case we consider.} What distinguishes them is their non-analytic dependence on the energy at the scale $E_b$.

The value of $\lambda_\ell(\mu_0)$ is fixed only up to its finite local part, which is matching data, determined by embedding the Schwarzian description into the complete four-dimensional quantum effective theory. The monopole is special: there is no running, $\text{B}_0=0$, and the shift symmetry $\Phi\to\Phi+c$ enforces $\lambda_0=0$ (see Section~\ref{sec:QSLN}).

\paragraph{Dissipative response: gap and threshold behavior.}
The imaginary part of the retarded correlator is finite, insensitive to local counterterms, and directly related to the absorption probability. It determines the dissipative Love numbers $\nu_\ell$, defined by the coefficient of $\beta\omega$ in $\operatorname{Im}\hat\chi_\ell(\omega)$ whenever the low-frequency expansion is regular. Here $\beta=2\pi/\sqrt{2E_bE}$ is the inverse temperature that the classical black hole of energy $E$ would have; in the microcanonical state it is only a label. Unlike its static counterpart, the monopole dissipative Love number does not vanish,
\begin{equation}
\label{eq:nu0_intro}
\nu_0 = \frac{r_+ E_b}{4\pi}\left[\sqrt{\frac{2E}{E_b}}\coth\left(2\pi\sqrt{\frac{2E}{E_b}}\right)+\frac{1}{2\pi}\right],
\end{equation}
which reduces to the classical value $r_+/(2\beta)$ for $E\gg E_b$. For $\ell\geq1$, instead, the rotational gap makes the dissipative response vanish near $\omega=0$ whenever the energy lies below the threshold of the spin-$\ell$ sector,
\begin{equation}
\nu_\ell = 0 \qquad \text{for} \quad E<E_{0,\ell}.
\end{equation}
This is the low-frequency signature of the quantum transparency of ref.~\cite{Emparan:2025qqf}. As $E$ approaches $E_{0,\ell}$ from above, the branch points of $\hat\chi_\ell(\omega)$ at
\begin{equation}
\omega=\pm(E-E_{0,\ell})
\end{equation}
approach the origin and the ordinary low-frequency expansion breaks down: the ratio $\nu_\ell/\nu_\ell^{\rm cl}$ diverges as $(E-E_{0,\ell})^{-1/2}$. At the threshold the Schwarzian continuum gives
\begin{equation}
\operatorname{Im}\hat\chi_\ell(\omega)
\propto\operatorname{sign}(\omega)\sqrt{|\omega|},
\end{equation}
and the corresponding branch cut produces a late-time $t^{-3/2}$ tail in the time-domain response. The strict threshold limit eventually probes energy scales where the smooth Schwarzian continuum approximation itself ceases to apply.

\paragraph{Plan.} In Section~\ref{sec:CLN}, we develop the classical response and relate the radial coefficient extraction to the renormalized two-point function. In Section~\ref{sec:QLN_def}, we compute and renormalize the Schwarzian-dressed retarded response. In Section~\ref{sec:QSLN}, we obtain the Love numbers: the static ones and their running, and the dissipative ones with their connection to absorption. We conclude in Section~\ref{sec:conclusions} with a discussion.

\bigskip

\noindent\emph{Note added.}
While this manuscript was being completed, ref.~\cite{Bhattacharjee:2026nmo} appeared, independently studying the static scalar response of near-extremal Reissner--Nordstr\"om black holes and defining the static Love numbers using a worldline effective field theory approach. Its results overlap with our static conservative sector and agree with ours up to overall normalization and local terms. Our approach is rather different, as we formulate the response directly in terms of the Schwarzian-dressed retarded Green's function. We analyze both its real and imaginary parts, including their full frequency dependence. This allows us to also study the dissipative sector, spectral thresholds, and branch-cut structure.

\section{Classical response function and Love numbers}
\label{sec:CLN}

We first define the response of the black hole to a scalar probe in the conventional way, from the solution of the wave equation, and compute it for a near-extremal black hole at leading order in $r_+\omega$. We then rewrite the result as a retarded two-point function at the mouth of the throat. Only the second formulation is valid when the throat fluctuates.

\subsection{Geometry, scales and radial regions}

The Reissner--Nordstr\"{o}m metric is
\begin{equation} \label{eq:RNfullgeometry}
    ds^2 = - f(r) \, dt^2 + \frac{dr^2 }{f(r)} + r^2 \left(d\theta^2 + \sin^2 \theta \, d\phi^2\right) \, , \qquad f(r) = \frac{(r - r_+) \, (r - r_-)}{r^2} \, ,
\end{equation}
with $r_\pm = M \pm \sqrt{M^2 - Q^2}$, in units $G = \hbar = 1$. Its inverse temperature is
\begin{equation}
\label{eq:beta_def}
    \beta = \frac{4 \pi r_+^2}{r_+ - r_-} \, ,
\end{equation}
and we measure the distance to extremality with
\begin{equation}
\label{eq:xi}
    \varepsilon \equiv \frac{r_+ - r_-}{r_+} = \frac{4 \pi r_+}{\beta} \, .
\end{equation}
The probe is a minimally coupled, neutral, massless scalar with action
\begin{equation}
\label{eq:scalaraction}
    I_\text{bulk} = -\frac{1}{2} \, \int d^4 x \, \sqrt{-g} \, \nabla_\mu \Phi \, \nabla^\mu \Phi \, .
\end{equation}
For a mode
\begin{equation}
\label{eq:Phi_ansatz}
    \Phi(t,r,\theta,\phi) = e^{-i \omega t} \, \psi(r) \, Y_{\ell m}(\theta, \phi) \, ,
\end{equation}
the radial function obeys
\begin{equation}
\label{eq:radialmaster}
   \partial_r(\Delta_r \, \psi') + \left[\frac{\omega^2 r^4}{\Delta_r} - \ell \, (\ell + 1) \right] \, \psi = 0 \, , \qquad \Delta_r = (r - r_+) \, (r - r_-) \, ,
\end{equation}
which does not depend on $m$.

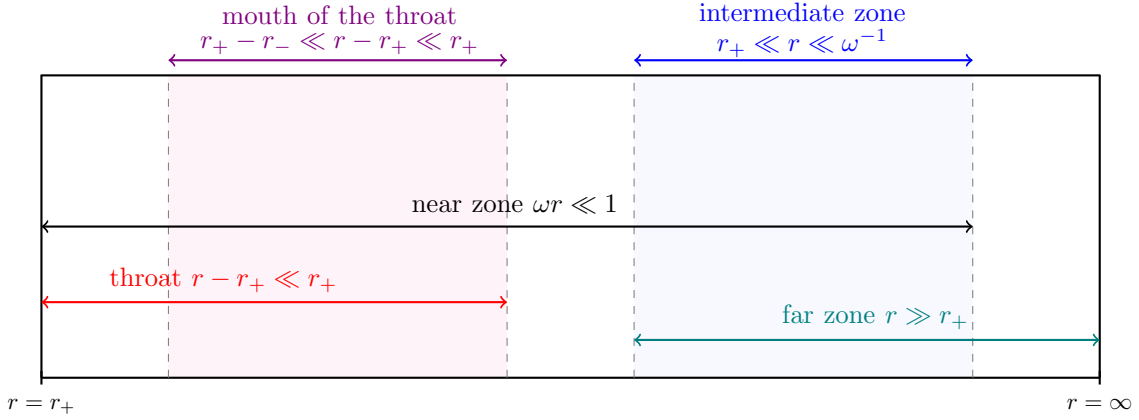
\begin{figure}
    \tikzstyle{xline}=[black,thick]
    \def\tick#1#2{\draw[thick] (#1) ++ (#2:0.09) --++ (#2-180:0.18)}
    \centering
    \def\xmax{14}
    \def\ymax{4}
    \begin{tikzpicture}
      \def\e{0.0*\xmax}
      \def\a{0.12*\xmax}
      \def\b{0.44*\xmax}
      \def\c{0.56*\xmax}
      \def\d{0.88*\xmax}
      \def\f{1.0*\xmax}
      \def\k{1.0*\ymax}
      \fill[blue,opacity=0.03] (\c,0) rectangle (\d,\k);
      \fill[magenta,opacity=0.05] (\a,0) rectangle (\b,\k);
      \draw[-,thick] (0,0) -- (0,\ymax) node[left] {};
      \draw[-,thick] (\xmax,0) -- (\xmax,\ymax) node[left] {};
      \draw[-,thick] (0,0) -- (\xmax,0) node[right=1,below] {};
      \draw[xline,line cap=round] (0,\k) -- (1.0*\xmax,\k) node[black,left=7,above=0,scale=0.9] {};
      \draw[dashed,gray] (\a,0) --++ (0,1.0*\k);
      \draw[dashed,gray] (\b,0) --++ (0,1.0*\k);
      \draw[dashed,gray] (\c,0) --++ (0,1.0*\k);
      \draw[dashed,gray] (\d,0) --++ (0,1.0*\k);
      \tick{\e,0}{90} node[below=-2.5,scale=0.8] {\strut$r=r_+$};
      \tick{\f,0}{90} node[below=-2.5,scale=0.8] {\strut$r=\infty$};
      \draw[<->,thick,red] (\e,1) -- (\b,1) node[above,text width=3cm,anchor=center] at (2.4,1.3) {\small \shortstack{throat $r-r_+ \ll r_+$} };
      \draw[<->,thick,teal] (\c,0.5) -- (\f,0.5) node[above,text width=3cm,anchor=center] at (11.3,0.8) {\small \shortstack{far zone $r \gg r_+$} };
      \draw[<->,thick,black] (\e,2) -- (\d,2) node[above,text width=3cm,anchor=center] at (6.4,2.3) {\small \shortstack{near zone $\omega r\ll 1$} };
      \draw[<->,thick,blue] (\c,4.2) -- (\d,4.2) node[above,text width=3cm,anchor=center] at (10.2,4.6) {\small \shortstack{intermediate zone \\ $r_+ \ll r \ll \omega^{-1}$} };
      \draw[<->,thick,violet] (\a,4.2) -- (\b,4.2) node[above,text width=3cm,anchor=center] at (3.6,4.6) {\small \shortstack{mouth of  the throat \\ $r_+-r_-\ll r-r_+\ll r_+$} };
    \end{tikzpicture}
    \caption{\small 
    Radial regions for a near-extremal Reissner--Nordstr\"{o}m black hole in the low-frequency regime $r_+\omega\ll1$. The near-zone is  $\omega r \ll 1$ while the far-zone is  $r\gg r_+$. Their overlap region is the (blue) intermediate zone $r_+ \ll r \ll \omega^{-1}$, where the response function can be read off. For near-extremal black holes, we can also define the near-horizon geometry, or throat, $r-r_+ \ll r_+$. The mouth of the throat (purple) is the asymptotic region of the near-horizon region: $r_+-r_- \ll r-r_+ \ll r_+$. The mouth region and the intermediate zone do not overlap directly, being separated by the (white) transition region $r-r_+ \sim \mathcal{O}(r_+)$. However, at leading order in $r_+\omega$, the solution \eqref{eq:u_NZ} remains valid through this transition region and the response functions computed from either the mouth of the throat or the intermediate zone are equivalent.
    }
    \label{fig:regions}
\end{figure}

We work at low frequency, $r_+ \omega \ll 1$, where the equation is solved by matching (Figure~\ref{fig:regions}). In the near zone, $\omega r \ll 1$, the frequency matters only close to the horizon, where it is enhanced by $1/\Delta_r$. In the far zone, $r \gg r_+$, the field propagates in almost flat space. The two overlap in the intermediate zone
\begin{equation}
\label{eq:IZ_region}
    \text{intermediate zone:} \qquad r_+ \ll r \ll \omega^{-1} \, ,
\end{equation}
where the solution is a sum of the two static power laws,
\begin{equation}
\label{eq:IZ_sol}
    \psi_\text{\tiny IZ}(r) = A \, \left(\frac{r}{r_+} \right)^\ell \, (1 + \cdots) + B \, \left(\frac{r}{r_+} \right)^{-(\ell + 1)} \, (1 + \cdots) \, ,
\end{equation}
with the dots standing for higher orders in $r_+/r$ and $\omega r$. The growing branch is the applied tidal field and the decaying one is the induced $\ell$-multipole moment. Their ratio is fixed by the boundary condition at the horizon and defines the response function,
\begin{equation}
\label{eq:chi_def}
    \hat{\chi}_\ell^\text{cl}(\omega) \equiv \frac{B}{2 A} \, ,
\end{equation}
where the factor $1/2$ follows the Newtonian convention \cite{Binnington:2009bb} and ``cl'' stands for classical.

At integer $\ell$ this definition is ambiguous: the subleading terms of the growing branch reach the order $r^{-\ell-1}$ of the decaying one, and a redefinition of $r$ reshuffles them. We resolve it in the usual way, by continuing $\ell$ to generic real values, for which the two series do not overlap and $B/A$ is invariant under redefinitions $r\to r\,(1+O(r_+/r))$, and returning to integer $\ell$ at the end \cite{Hui:2020xxx,LeTiec:2020bos,Cano:2025zyk}. Matching to a scattering amplitude, or to a Wilson coefficient of the worldline effective theory, gives the same quantity an observable meaning \cite{Ivanov:2022hlo}.\footnote{At higher orders in $r_+\omega$ the continuation in $\ell$ requires more care \cite{Chakraborty:2025wvs}, which we do not need here.}

At $\omega=0$ the response can be computed exactly for any charge and in any dimension. We do this in Appendix~\ref{app:CLN_Ddim}; in four dimensions it coincides with the static limit of the result in \eqref{eq:chi_result}.

\subsection{The classical response of the near-extremal throat}

It is convenient to trade $r$ and $t$ for
\begin{equation}
\label{eq:Z_def}
    Z = \frac{r_+ - r_-}{2 r - r_+ - r_-} \, , \qquad \tau = \frac{2\pi t}{\beta} \, .
\end{equation}
The coordinate $Z$ decreases from $Z=1$ at the outer horizon to $Z=0$ at $r\to\infty$, the inner horizon is at $Z=-1$, and $\Delta_r = (r_+ - r_-)^2 (1-Z^2)/(4Z^2)$ exactly. Throughout the near zone, we can set $r^4 \to r_+^4$ in the frequency term of \eqref{eq:radialmaster}: at the horizon this is exact, and away from it the term is negligible.\footnote{The term neglected in \eqref{eq:radialmaster_scaling} is $\tilde\omega^2 \left[(r/r_+)^4 - 1 \right] \psi/(1-Z^2)$. For $r - r_+ \ll r_+$ it equals $2 \varepsilon \tilde\omega^2 \psi/[Z(1+Z)]$. Compared to the terms kept, it is of order $\varepsilon\tilde\omega^2 = r_+\omega\, \beta\omega/\pi$ along the throat, down to the horizon. For $r\gg r_+$, it is of relative order $(\omega r)^2$. For $r \gg r_+$, the near-horizon approximation is no longer valid, and one must instead include the complete term $\tilde{\omega}^{2}[(r/r_+)^{4}-1]\psi/(1-Z^{2})$, which is of relative order $(\omega r)^2$.} The radial equation then becomes
\begin{equation}
\label{eq:radialmaster_scaling}
   \partial_Z \! \left[ (1-Z^2) \, \partial_Z \psi \right] + \left[\frac{\tilde{\omega}^2}{1-Z^2} - \frac{\ell \, (\ell + 1)}{Z^2} \right] \, \psi = 0 \, , \qquad \tilde{\omega} \equiv \frac{\beta \omega}{2 \pi} \, ,
\end{equation}
which is Legendre's equation in the variable $1/Z$. So far the charge is arbitrary as we have not yet specialized to near extremality. To do so, we note that near extremality, the temperature is low, and we can take
\begin{equation}
\label{eq:NE_scaling}
    \varepsilon \ll 1 \, , \qquad \tilde{\omega} = O(1) \, , \qquad \text{hence} \quad r_+ \omega = \frac{\varepsilon \tilde{\omega}}{2} \ll 1 \, ,
\end{equation}
with the hierarchy $r_+ - r_- \ll r_+ \ll \omega^{-1}$. All the results of this section are at leading order in $\varepsilon$ with $\tilde\omega$ fixed: leading order in $r_+\omega$, exact in $\beta\omega$. The black hole is classical, which requires $1/\beta$ to be small compared to $1/r_+$ but large compared to the scale $E_b$ of \eqref{eq:Eb}.

In the same limit, $\varepsilon\to 0$ at fixed $Z$ and $\tau$, the geometry \eqref{eq:RNfullgeometry} becomes
\begin{equation}
\label{eq:NH_metric_Z}
    ds^2 = \frac{r_+^2}{Z^2} \left[ -(1-Z^2) \, d\tau^2 + \frac{dZ^2}{1-Z^2} \right] + r_+^2 \left( d\theta^2 + \sin^2 \theta \, d\phi^2 \right) \, .
\end{equation}
Thus the near-horizon near-extremal geometry of the black hole is given by $\text{AdS}_2$ times a sphere, both of radius $r_+$. The Euclidean section is smooth when $\tau\sim\tau+2\pi$, that is, $t\sim t+\beta$. Equation \eqref{eq:radialmaster_scaling} is the wave equation in this geometry, with $\tilde\omega$ being the frequency conjugate to $\tau$. The throat \eqref{eq:NH_metric_Z} extends over $r - r_+ \ll r_+$, that is, $Z \gg \varepsilon$, and when $\varepsilon\ll 1$ this leaves room for an asymptotic region inside it, which we call its mouth,
\begin{equation}
\label{eq:mouth_region}
    \text{mouth of the throat:} \qquad r_+-r_-\ll r-r_+\ll r_+ \qquad \Longleftrightarrow \qquad \varepsilon \ll Z \ll 1 \, .
\end{equation}
There, the temperature is negligible and the metric is that of the extremal throat, $r_+^2(-dt^2+d\zeta^2)/\zeta^2$, with
\begin{equation}
\label{eq:zeta_def}
    \zeta \equiv \frac{\beta Z}{2\pi} = \frac{2 r_+ Z}{\varepsilon} \simeq \frac{r_+^2}{r-r_+} \, .
\end{equation}
The mouth does not overlap with the intermediate zone \eqref{eq:IZ_region}, which lies at $Z \ll \varepsilon$, but \eqref{eq:radialmaster_scaling} holds across both, since its validity is that of the near zone and not only of the throat. The general solution of \eqref{eq:radialmaster_scaling} is
\begin{equation}
\label{eq:u_NZ}
    \psi_\text{\tiny NZ} = c_1 \, P_\ell^{i \tilde{\omega}}(1/Z) + c_2 \, Q_\ell^{i \tilde{\omega}}(1/Z) \, ,
\end{equation}
in terms of associated Legendre functions. The retarded response requires the solution to be ingoing at the future horizon \cite{Son:2002sd}, $\psi \sim e^{-i\omega r_\ast} \sim (1-Z)^{-i\tilde\omega/2}$. This is the behavior of $P_\ell^{i\tilde\omega}$ and not of $Q_\ell^{i\tilde\omega}$, so $c_2 = 0$. At small $Z$,
\begin{equation}
\label{eq:u_NH_Zto0}
    P_\ell^{i \tilde{\omega}}(1/Z) = \frac{\Gamma\!\left(\ell + \frac{1}{2} \right)}{\sqrt{\pi} \, \Gamma(\ell + 1 - i \tilde{\omega})} \, \left( \frac{Z}{2} \right)^{-\ell} \left[1 + O(Z^{2}) \right] + \frac{\Gamma\!\left(-\ell - \frac{1}{2} \right)}{\sqrt{\pi} \, \Gamma(-\ell - i \tilde{\omega})} \, \left( \frac{Z}{2} \right)^{\ell+1} \left[1 + O(Z^{2}) \right] \, .
\end{equation}
In the intermediate zone $Z \simeq \varepsilon r_+/(2r)$, and comparing with \eqref{eq:IZ_sol} we find
\begin{equation}
\label{eq:AB_explicit}
    A = c_1 \left(\frac{4}{\varepsilon} \right)^{\ell} \frac{\Gamma\!\left(\ell + \frac{1}{2} \right)}{\sqrt{\pi} \, \Gamma(\ell + 1 - i \tilde{\omega})} \, , \qquad
    B = c_1 \left(\frac{\varepsilon}{4} \right)^{\ell+1} \frac{\Gamma\!\left(-\ell - \frac{1}{2} \right)}{\sqrt{\pi} \, \Gamma(-\ell - i \tilde{\omega})} \, ,
\end{equation}
so that, with $\varepsilon/4 = \pi r_+/\beta$,
\begin{equation}
\label{eq:chi_result}
    \hat{\chi}_\ell^\text{cl}(\omega) = \frac{1}{2} \left(\frac{\pi r_+}{\beta} \right)^{2 \ell + 1} \frac{\Gamma\!\left(-\ell - \frac{1}{2} \right) \, \Gamma\!\left(\ell + 1 - \frac{i \beta \omega}{2 \pi} \right)}{\Gamma\!\left(\ell + \frac{1}{2} \right) \, \Gamma\!\left(-\ell - \frac{i \beta \omega}{2 \pi} \right)} \, .
\end{equation}
The same ratio is obtained at the mouth, where $Z \simeq \varepsilon r_+/[2(r-r_+)]$, with $(r-r_+)/r_+$ in place of $r/r_+$ in \eqref{eq:IZ_sol}. In this sense the response of the black hole is, at leading order in $r_+\omega$, the response of the throat measured at its mouth. This is the quantity that the Schwarzian mode will correct in Section~\ref{sec:QLN_def}.

The Love numbers are the coefficients of the low-frequency expansion of the response \cite{Ivanov:2022hlo}, which in the regime \eqref{eq:NE_scaling} is an expansion in $\beta\omega$,\footnote{The Love numbers are usually defined by a perturbative expansion in $r_+\omega$ of the response in the intermediate zone. However, we work in leading order $r_+\omega$ and so we neglect the subleading corrections in this expansion.} 
\begin{equation}
\label{eq:chi_series_LN}
    \hat{\chi}^\text{cl}_\ell(\omega) = \lambda^\text{cl}_\ell + i \, \nu^\text{cl}_\ell \, \beta \omega + O\!\left((\beta \omega)^2 \right) \, ,
\end{equation}
where $\lambda^\text{cl}_\ell$ is the static Love number and $\nu^\text{cl}_\ell\geq 0$ is the dissipative Love number. For real $\ell$, the reflection formula of the Gamma function gives
\begin{align}
    \label{eq:Re_chiFT_cl}
    \text{Re}\,\hat{\chi}^\text{cl}_\ell(\omega) &= \frac{1}{2} \, \left(\!\frac{\pi r_+}{\beta} \!\right)^{2 \ell + 1} \, \frac{\left|\Gamma\!\left(\ell + 1 + \frac{i \beta \omega}{2 \pi} \right) \right|^2}{\left(\ell + \frac{1}{2} \right) \, \Gamma\!\left(\ell + \frac{1}{2} \right)^2} \, \tan(\pi \ell) \, \cosh\!\left(\frac{\beta \omega}{2} \right) \, ,\\
    \label{eq:Im_chiFT_cl}
    \text{Im}\,\hat{\chi}^\text{cl}_\ell(\omega) &= \frac{1}{2} \, \left(\!\frac{\pi r_+}{\beta} \!\right)^{2 \ell + 1} \, \frac{\left|\Gamma\!\left(\ell + 1 + \frac{i \beta \omega}{2 \pi} \right) \right|^2}{\left(\ell + \frac{1}{2} \right) \, \Gamma\!\left(\ell + \frac{1}{2} \right)^2} \, \sinh\!\left(\frac{\beta \omega}{2} \right) \, ,
\end{align}
and therefore
\begin{align}
    \label{eq:CSLN_def}
    \lambda^\text{cl}_\ell &= \lim_{\omega \to 0} \text{Re}\,\hat{\chi}^\text{cl}_\ell(\omega)
    =\frac{1}{2} \, \left(\!\frac{\pi r_+}{\beta} \!\right)^{2 \ell + 1} \, \frac{\Gamma\!\left(\ell + 1 \right)^2}{\left(\ell + \frac{1}{2} \right) \, \Gamma\!\left(\ell + \frac{1}{2} \right)^2} \, \tan(\pi \ell) \, ,\\
    \label{eq:CDLN_def}
    \nu^\text{cl}_\ell &= \lim_{\omega \to 0} \frac{\text{Im}\,\hat{\chi}^\text{cl}_\ell(\omega)}{\beta \omega}
    = \frac{1}{4} \, \left(\frac{\pi r_+}{\beta} \right)^{2 \ell + 1} \, \frac{\Gamma(\ell + 1)^2}{\left(\ell + \frac{1}{2} \right) \, \Gamma\!\left(\ell + \frac{1}{2} \right)^2} \, .
\end{align}
After translating conventions, these agree with the four-dimensional results of \cite{Charalambous:2024tdj}, and \eqref{eq:CSLN_def} with the exact static result \eqref{eq:lambda_cl_RN} of Appendix~\ref{app:CLN_Ddim}. We analytically continue to integer $\ell$, where the factor $\tan(\pi\ell)$ vanishes, and with it the whole conservative response,
\begin{equation}
\label{eq:Re_chi_int}
    \text{Re}\,\hat{\chi}^\text{cl}_\ell(\omega) = 0 \, , \qquad \ell \in \mathbb{N} \, ,
\end{equation}
so in particular
\begin{equation}
\label{eq:lambda_cl_zero}
    \lambda^\text{cl}_\ell = 0 \, , \qquad \ell \in \mathbb{N} \, .
\end{equation}
This holds to all orders in $\beta\omega$ at leading order in $r_+\omega$. However, at higher orders in $r_+\omega$ the conservative dynamical coefficients do not necessarily vanish \cite{Chakraborty:2025wvs}. For the imaginary part, taking integer $\ell$ in \eqref{eq:Im_chiFT_cl} leads to
\begin{equation}
\label{eq:Im_chi_int}
    \text{Im}\,\hat{\chi}^\text{cl}_\ell(\omega) = \frac{\pi}{2} \, \left(\!\frac{\pi r_+}{\beta} \!\right)^{2 \ell + 1} \, \frac{1}{\left(\ell + \frac{1}{2} \right) \, \Gamma\!\left(\ell + \frac{1}{2} \right)^2} \, \frac{\beta \omega}{2 \pi} \, \prod_{n=1}^\ell \left[n^2 + \left(\frac{\beta \omega}{2 \pi} \right)^2 \right] \, , \qquad \ell \in \mathbb{N} \, ,
\end{equation}
which is in general nonzero. The dissipative Love numbers for the first few integer values of $\ell$ are found to be
\begin{equation}
\label{eq:nu_cl}
    \nu^\text{cl}_0 = \frac{r_+}{2 \beta} \, , \qquad \nu^\text{cl}_1 = \frac{2}{3 \pi} \, \left(\!\frac{\pi r_+}{\beta} \!\right)^3 \, , \qquad \nu^\text{cl}_2 = \frac{32}{45 \pi} \, \left(\!\frac{\pi r_+}{\beta} \!\right)^5 \, , \qquad \nu^\text{cl}_3 = \frac{128}{175 \pi} \, \left(\!\frac{\pi r_+}{\beta} \!\right)^7 \, .
\end{equation}
These are the classical values to which we will compare the quantum ones in Section~\ref{sec:QDLN}.

\subsection{Retarded two-point function and holographic renormalization} \label{subsec:retarted2pointfunction}

The response \eqref{eq:chi_result} is a property of the throat, and it can be phrased in the language of holography. Restoring the sum over partial waves,
\begin{equation}
    \Phi = \sum_{\ell,m} \Psi_{\ell m}(t,Z) \, Y_{\ell m}(\theta,\phi) \, , \qquad \Psi_{\ell m}^\ast = (-1)^m \, \Psi_{\ell, -m} \, ,
\end{equation}
the action \eqref{eq:scalaraction} in the throat reduces to
\begin{equation}
\label{eq:I_bulk_cl_aux}
    I_\text{bulk} = -\frac{1}{2} \, \int d^2 x \, \sqrt{-g_2} \, \sum_{\ell,m} \left[r_+^2 \, \nabla_a \Psi_{\ell m}^\ast \, \nabla^a \Psi_{\ell m} + \ell (\ell + 1) \, \Psi_{\ell m}^\ast \, \Psi_{\ell m} \right] \, ,
\end{equation}
where $g_2$ is the $\text{AdS}_2$ factor of \eqref{eq:NH_metric_Z}. Each partial wave is a scalar in $\text{AdS}_2$ with mass and conformal dimension
\begin{equation}
\label{eq:Delta}
    m_\ell^2 r_+^2 = \ell (\ell + 1) = \Delta(\Delta-1) \, , \qquad \Delta = \ell + 1 \, .
\end{equation}
Towards the boundary it behaves as in \eqref{eq:u_NH_Zto0},
\begin{equation}
\label{eq:hat_psi_A_B}
    \Psi_{\ell m}(t,Z) = A_{\ell m}(t) \, \left(\frac{\zeta}{r_+} \right)^{1-\Delta} \left[1 + \cdots \right] + B_{\ell m}(t) \, \left(\frac{\zeta}{r_+} \right)^{\Delta} \left[1 + \cdots \right] \, ,
\end{equation}
with $\zeta$ given in \eqref{eq:zeta_def}. 
The coefficients $A_{\ell m}$ and $B_{\ell m}$ are those of \eqref{eq:IZ_sol}: in the mouth $\zeta/r_+ = r_+/(r-r_+)$, so the two power laws are written in terms of $(r-r_+)$ in units of $r_+$, with no factors of the temperature.
For the ingoing solution, the Fourier components, $A_{\ell m}(t) = \int d\omega \, e^{-i\omega t} A_{\ell m}(\omega)$, are related by $B_{\ell m}(\omega) = 2 \hat\chi^\text{cl}_\ell(\omega) A_{\ell m}(\omega)$.

In the quantum mechanics dual to the throat, $A_{\ell m}(t)$ is the source of an operator $O_{\ell m}(t)$ of dimension $\Delta$,
\begin{equation}
\label{eq:I_source_cl}
    I_\text{source} = \int dt \, \sum_{\ell,m} A_{\ell m}(t) \, O_{\ell m}^\dagger(t) \, , \qquad O_{\ell m}^\dagger = (-1)^m O_{\ell,-m} \, .
\end{equation}
To linear order, the source induces the expectation value
\begin{equation}
\label{eq:lin_resp}
    \left\langle O_{\ell m}(t) \right\rangle = \int dt' \, \left\langle O_{\ell m}(t) O_{\ell m}^\dagger(t') \right\rangle_R \, A_{\ell m}(t') \, ,
\end{equation}
where
\begin{equation}
\label{eq:th_ret_corr_def}
    \left\langle O_{\ell m}(t) O_{\ell m}^\dagger(t') \right\rangle_R
    = i \, \Theta(t-t') \, \left\langle \left[ O_{\ell m}(t) , O_{\ell m}^\dagger(t') \right] \right\rangle_\beta
\end{equation}
is the thermal retarded two-point function. Holographically, the expectation value is the renormalized radial momentum conjugate to the source \cite{Iqbal:2009fd}. We cut off the action at a surface $Z = Z_0$ in the mouth and add counterterms $I_\text{ct}$: local functionals of the field on that surface, with coefficients that do not depend on the state, which cancel the terms that diverge as $Z_0\to 0$ \cite{deHaro:2000vlm,Skenderis:2002wp}. Under a change of the source, the action evaluated on the solution that is ingoing at the horizon \cite{Son:2002sd,Skenderis:2008dg} then varies at the cutoff by
\begin{equation} \label{eq:deltaIbulkct}
    \delta \left( I_\text{bulk} + I_\text{ct} \right) \Big|_{Z_0} \; \xrightarrow[Z_0\to 0]{} \; \sum_{\ell,m} \int dt \, \left\langle O_{\ell m}(t) \right\rangle \, \delta A_{\ell m}^\ast(t) \, .
\end{equation}
The result is the standard one for a scalar in $\text{AdS}_{d+1}$, here with $d=1$ \cite{Balasubramanian:1998de,Klebanov:1999tb,deHaro:2000vlm}:
\begin{equation}
\label{eq:O_vev}
    \left\langle O_{\ell m}(t) \right\rangle = (2\Delta - 1) \, r_+ \, B_{\ell m}(t) \, .
\end{equation}
The renormalized one-point function is the coefficient of the decaying branch. With $\hat{O}_{\ell m}(\omega) = \int dt \, e^{i\omega t} O_{\ell m}(t)$ and $\hat{O}_{\ell m}^\dagger(\omega) \equiv ( \hat{O}_{\ell m}(\omega) )^\dagger$, eqs.~\eqref{eq:lin_resp} and \eqref{eq:O_vev} give
\begin{equation}
\label{eq:2pt_R}
    \left\langle \hat{O}_{\ell m}(\omega)
    \hat{O}_{\ell m}^\dagger(\omega') \right\rangle_R
    =
    4\pi r_+(2\ell+1) \,
    \delta(\omega-\omega') \,
    \hat{\chi}_\ell^\text{cl}(\omega) \, .
\end{equation}
Reading $B/A$ from the radial profile is thus the same as computing the retarded correlator from the renormalized generating functional, and the response function is this correlator up to a kinematic normalization.\footnote{The retarded prescription varies the renormalized radial momentum of the ingoing solution \cite{Son:2002sd,Skenderis:2008dg}.
Ordinary second differentiation of the real on-shell quadratic expression retains only the conservative part of the response.}

In general \eqref{eq:O_vev} holds up to finite local functionals of the source, which reflect the freedom to add finite counterterms and make the contact terms of the correlator depend on the scheme. Here there are none, and for a single reason: $2\Delta - 1 = 2\ell+1$ is odd.

First, there are no logarithms. As we saw, both series in \eqref{eq:u_NH_Zto0} proceed in powers of $Z^{2}$ and their leading terms differ by $Z^{2\Delta-1}$. A logarithmic term, and with it a renormalization scale and a scheme-dependent finite part, appears only when the two series overlap, that is, when $2\Delta - 1$ is an even integer \cite{deHaro:2000vlm,Skenderis:2002wp}.

Second, there are no finite local terms. A counterterm is local in the source, with a coefficient that does not depend on the state. The state can enter it only through an operator of the boundary theory, and for a non-rotating black hole the only one available is the Hamiltonian $H$. It appears through the Schwarzian derivative of the boundary time, that is, in the combination $E_b H$ --- where $E_{b}$ is defined in \eqref{eq:Eb} --- which has dimension two and takes the value $E_b E = 2\pi^2/\beta^2$ on the classical black hole, cf.~\eqref{eq:beta_NE}. As shown in Appendix~\ref{app:ct}, the combination $E_bE$ is related to the extrinsic curvature $K$ of the cutoff surface. A local term quadratic in the source is then of the form
\begin{equation}
\label{eq:source_alpha}
    \sum_{m=-\ell}^{\ell} \int dt \, A_{\ell m}^\ast(t) \, \left( E_b H \right)^c \, \partial_t^\alpha A_{\ell m}(t) \, , \qquad c = 0, 1, 2, \ldots \, ,
\end{equation}
with a numerical coefficient times a power of the cutoff. For $\ell\geq 1$ the source is irrelevant, $\Delta > 1$, and divergent terms of this kind with $c \geq 1$ do occur; we show this in the bulk for $\ell=1$ in Appendix~\ref{app:ct}. Since $A_{\ell m}$ has dimension $1-\Delta = -\ell$, the term is independent of the cutoff only if $\alpha + 2c = 2\Delta - 1 = 2\ell+1$, so that $\alpha$ is odd,
\begin{equation}
\label{eq:J_odd}
    \left( \frac{2\pi^2}{\beta^2} \right)^{\! c} \, \sum_{m=-\ell}^{\ell} \int dt \, A_{\ell m}^\ast(t) \, \partial_t^{2(\ell-c)+1} A_{\ell m}(t) \, .
\end{equation}
Using $A_{\ell m}^\ast = (-1)^m A_{\ell,-m}$ and integrating by parts an odd number of times, this is equal to minus itself, and vanishes. In other words, a local term would contribute to $\hat\chi_\ell$ a real polynomial in $\omega^2$ and $E_b E$ of overall energy dimension $(2\ell+1)$. Since both $\omega^2$ and $E_b E$ have dimension two, no such polynomial exists, in agreement with \eqref{eq:Re_chi_int}.

The classical response is therefore free of renormalization ambiguities. The argument rests on the absence of any scale of odd dimension. In the quantum theory the loops of the Schwarzian mode bring powers of $E_b$ that are not accompanied by $H$. Since $E_b$ has dimension one, both logarithms and finite local terms then appear, as we show in Section~\ref{sec:renorm}.

\section{Quantum response function}
\label{sec:QLN_def}

In Section~\ref{sec:CLN} we computed the classical response function at the mouth of the throat of a near-extremal black hole by leveraging the fact that the near-horizon region becomes an $\text{AdS}_2$ geometry. However, as the black hole gets closer to extremality, the throat undergoes large quantum fluctuations, which drastically modify semiclassical predictions \cite{Iliesiu:2020qvm}. In light of this, it is crucial to revisit our classical response function to account for quantum effects. These fluctuations are captured by the Schwarzian theory, and become important when the black hole energy above extremality, $E \equiv M-Q$, is of order
\begin{equation}
\label{eq:Eb}
    E_b = \frac{\pi}{r_+ S_0} \, , \qquad S_0 = \frac{\pi r_+^2}{G}\,,
\end{equation}
where $S_0$ denotes the (naive) semiclassical entropy at extremality. The characteristic scale $E_b$ is the energy at which the semiclassical description of gravity breaks down. Therefore, we are in the quantum regime once we are close to this energy scale:
\begin{equation}
\label{eq:quantum_reg}
    \text{quantum regime:} \qquad E \lesssim E_b\,.
\end{equation}
The Schwarzian description holds for macroscopic black holes near extremality, $S_0\gg1$ and $\varepsilon\ll1$ \cite{Iliesiu:2020qvm}. Since $E\simeq r_+\varepsilon^2/8$, this means $E\ll S_0/r_+$. The scale $E_b\ll 1/r_+$ marks the onset of strong fluctuations, but the description also covers the range $E_b\ll E\ll S_0/r_+$, where the black hole is classical.

Quantum effects only modify the dynamics of the near-horizon region $r/r_{+}-1\ll 1$. 
Thus, even in the quantum regime \eqref{eq:quantum_reg}, the geometry outside the throat remains classical. Consequently, the propagation between the mouth of the throat \eqref{eq:mouth_region} and the intermediate zone \eqref{eq:IZ_region} described in Section~\ref{sec:CLN} is unchanged. 
In particular, this implies that at leading order in $\omega r_{+}$, the response measured at the mouth of the throat is the same as the response measured in the intermediate zone. 
Hence, it is sufficient to determine the quantum response function at the mouth of the throat, as this already represents the response in the intermediate zone at leading order in $r_+\omega$. We will neglect all subleading corrections in $r_+\omega$ due to propagation through the classical exterior --- these are unaffected by the Schwarzian mode and much smaller than the effects of order $\beta\omega$ which we do include.

The goal of this section is therefore to compute the quantum response function at the mouth of the throat --- \textit{i.e.} the quantum version of \eqref{eq:chi_result}. We achieve this by coupling a scalar source to the Schwarzian mode and computing the associated retarded Green's function, as we explain next.

\subsection{Setup}

We consider a fixed partial wave $\ell$ of the scalar field $\Phi(t,r,\theta,\phi)$, as in \eqref{eq:Phi_ansatz}. This perturbation acts as a source at the mouth of the $\text{AdS}_2$ throat. To distinguish the normalization convenient for the Schwarzian matrix elements from the bulk normalization used in Section~\ref{sec:CLN}, we denote the source and the corresponding boundary operator by $\varphi_{\ell m}(t)$ and $\mathcal{O}_{\ell m}(t)$, respectively. Their relative normalization is fixed below by matching to the classical response. The operator $\mathcal{O}_{\ell m}(t)$ has conformal dimension $\Delta=\ell+1$, as given in \eqref{eq:Delta}. In the standard complex spherical-harmonics basis, the reality of the source field and the Hermiticity of the response operator imply
\begin{equation}
\label{eq:reality_lm}
    \varphi_{\ell m}^\ast(t) = (-1)^m \, \varphi_{\ell,-m}(t) \, , \qquad \mathcal{O}_{\ell m}^\dagger(t) = (-1)^m \, \mathcal{O}_{\ell,-m}(t) \, .
\end{equation}
At low energies, the dynamics of the near-extremal black hole is described by the Schwarzian mode $I_\text{Sch}$ together with a rotational massless mode $I_{\text{SO}(3)}$ associated with the transverse $S^2$. The coupling of the scalar perturbation to the near-extremal black hole is therefore encoded, at low energies, by the boundary action
\begin{equation}
\label{eq:action}
    I = I_\text{Sch} + I_{\text{SO}(3)} + \sum_{m=-\ell}^{\ell} \int dt \, \varphi_{\ell m}(t) \, \mathcal{O}_{\ell m}^\dagger(t) \, .
\end{equation}
In particular, \eqref{eq:reality_lm} ensures that the action \eqref{eq:action} is real. As we will discuss below, in order to renormalize the Schwarzian-dressed conservative response, one also needs to add suitable counterterms to the action \eqref{eq:action}.

The response to the source $\varphi_{\ell m}$ is given by the retarded Green's function
\begin{equation}
\label{eq:GR_time_def}
    G_\ell^R(t) = i \, \Theta(t) \, \sum_{m=-\ell}^{\ell} \left\langle \left[\mathcal{O}_{\ell m}(t), \mathcal{O}_{\ell m}^\dagger(0) \right] \right\rangle \, , \qquad t \in \mathbb{R} \, .
\end{equation}
We are mostly interested in the frequency-domain response, given by its Fourier transform 
\begin{equation}
\label{eq:GR_FT_def}
    \widehat{G}_\ell^R(\omega) = \lim_{\eta \to 0^+} \int_{-\infty}^{+\infty} dt \, e^{i (\omega + i \eta) t} \, G^R_\ell(t) \, , \qquad \omega \in \mathbb{R} \, .
\end{equation}
The small positive imaginary part $i \eta$ defines the retarded Green's function at real frequency $\omega \in \mathbb{R}$ as the boundary value of a function analytic in the upper half-plane, $\text{Im}\,\omega > 0$. We define the dimensionless quantum response function by
\begin{equation}
\label{eq:chi_quantum}
    \hat{\chi}_\ell(\omega) = \mathcal{N}_\ell \, \widehat{G}_\ell^R(\omega) \, ,
\end{equation}
where $\mathcal{N}_{\ell}$ is a normalization constant fixed by requiring that the classical limit of \eqref{eq:chi_quantum} reproduces the classical response function
\eqref{eq:chi_result}.  Anticipating our results below, we find that this constant is given by 
\begin{equation}
\label{eq:sN_1}
\mathcal{N}_\ell
=
\frac{\Gamma(\ell+1)}
{4\sqrt{\pi}\,\Gamma\!\left(\ell+\frac32\right)}
\,r_+^{2\ell+1}.
\end{equation}
We then define the quantum static and dissipative Love numbers analogously to \eqref{eq:CSLN_def} and \eqref{eq:CDLN_def}, which gives us
\begin{equation}
\label{eq:QLN_def}
    \lambda_\ell = \lim_{\omega \to 0} \text{Re}\,\hat{\chi}_\ell(\omega) \, , \qquad \nu_\ell = \lim_{\omega \to 0} \frac{\text{Im}\,\hat{\chi}_\ell(\omega)}{\beta \omega} \, .
\end{equation}
In other words, we perform the same low-frequency expansion of the response function as in the classical case \eqref{eq:chi_series_LN}, which takes the form
\begin{equation}
\label{eq:chi_qu_series_LN}
    \hat{\chi}_\ell(\omega) = \lambda_\ell + i \, \nu_\ell \, \beta \omega + O\!\left((\beta \omega)^2 \right) \, .
\end{equation}
We evaluate these quantities in the microcanonical ensemble, in which the black hole energy and angular momentum are fixed. We denote by $j$ and $j'$ the spin of the black hole (in units of $\hbar$) before and after the perturbation $\varphi_{\ell m}(t)$, respectively, and by $m_\text{\tiny BH} \in \{-j,\ldots,j\}$ and $m'_\text{\tiny BH} \in \{-j',\ldots,j'\}$ the corresponding azimuthal mode numbers. Composition of angular momenta requires $|j-\ell| \leq j' \leq j+\ell$ and $m'_\text{\tiny BH} = m_\text{\tiny BH} + m$.

The semiclassical inverse temperature \eqref{eq:beta_def} is given, at leading order near extremality, by
\begin{equation}
\label{eq:beta_NE}
    \beta = \frac{2 \pi}{\sqrt{2 E_b E}} \, .
\end{equation}
For $\ell \ge 1$, the scalar perturbation carries angular momentum and can change the spin of the black hole. To account for this effect, besides the Schwarzian fluctuations of the $\text{AdS}_2$ region, we must also keep track of the spin sector associated with the transverse $S^2$. It is convenient to parametrize the energy in a spin-$j$ sector by
\begin{equation}
\label{eq:k_def}
    k = \sqrt{\frac{2 (E - E_{0,j})}{E_b}} \, ,
\end{equation}
where $E$ is the black hole energy above extremality, and
\begin{equation}
\label{eq:gap}
    E_{0,j} = \frac{j (j+1)}{2} \, E_b
\end{equation}
is a spin-induced energy gap in the quantum black hole spectrum. The density of black hole states at fixed $j$ and fixed $m_\text{\tiny BH}$ is given by \cite{Iliesiu:2020qvm,Brown:2024ajk}
\begin{align}
\label{eq:rho_j}
    \rho_j(E)
    &= \frac{(2 j + 1) e^{S_0}}{2 \pi^2 E_b} \, \sinh\!\left(2 \pi \sqrt{\frac{2 (E - E_{0,j})}{E_b}} \right) \, \Theta(E - E_{0,j}) \, .
\end{align}
The total density in the spin-$j$ sector contains an additional factor of $(2j+1)$ from the sum over $m_\text{\tiny BH}$. States in this sector are absent below the threshold $E = E_{0,j}$, which is taken into account via the step function in \eqref{eq:rho_j}.

To compute Green's functions, we also need the matrix elements of $\mathcal{O}_{\ell m}$ between spinning black hole states:
\begin{equation}
\label{eq:matrix_elements}
    \left| \langle E', j', m'_\text{\tiny BH} | \mathcal{O}_{\ell m} | E, j, m_\text{\tiny BH} \rangle \right|^2 = \frac{\left|C^{j' \, m'_\text{\tiny BH}}_{j \, m_\text{\tiny BH}, \, \ell \, m} \right|^2}{2 j' + 1} \, \frac{2 \, e^{-S_0}}{(2/E_b)^{2 \Delta} \, \Gamma(2 \Delta)} \, \gamma_\Delta(k,k') \, ,
\end{equation}
where we have defined
\begin{align}
    \gamma_\Delta(k,k') \equiv \Gamma(\Delta \pm i k \pm i k') \, , \qquad k' = \sqrt{\frac{2 (E' - E_{0,j'})}{E_b}} \, .
\end{align}
Here $\Gamma(\Delta \pm i k \pm i k')$ denotes the product of Gamma functions over the four independent choices of signs. Moreover, we denote the Clebsch--Gordan coefficients of $\text{SU}(2)$ by $C^{j' \, m'_\text{\tiny BH}}_{j \, m_\text{\tiny BH}, \, \ell \, m}$, which satisfy the identity
\begin{equation}
\label{eq:CG_identity}
    \sum_{m'_\text{\tiny BH}, m_\text{\tiny BH}, m} \left|C^{j' \, m'_\text{\tiny BH}}_{j \, m_\text{\tiny BH}, \, \ell \, m} \right|^2 = 2 j' + 1 \, .
\end{equation}
For simplicity, throughout the remainder of this work we restrict to an initially spinless black hole,
\begin{equation}
\label{eq:j0_1}
    j = 0 \, , \qquad m_\text{\tiny BH} = 0 \, ,
\end{equation}
for which
\begin{equation}
\label{eq:j0_2}
    j' = \ell \, , \qquad m'_\text{\tiny BH} = m \, .
\end{equation}
Before turning to the calculation of the quantum response function, let us discuss an additional symmetry that will play an important role below. The four-dimensional scalar action \eqref{eq:scalaraction} is invariant under the constant shift
\begin{equation}
\label{eq:scalar_shift}
    \Phi\to\Phi+c \, ,
    \qquad c=\text{constant} \, ,
\end{equation}
since the scalar enters the action only through its derivatives. Since Einstein--Maxwell theory is independent of $\Phi$, this is also a symmetry of the full classical system. In the spherical-harmonic and Fourier decompositions, the transformation \eqref{eq:scalar_shift} acts only on the static monopole, $\ell=0$ and $\omega=0$. In the low-energy description \eqref{eq:action}, it therefore corresponds to shifting the constant monopole source $\varphi_{00}$ by
\begin{equation}
\label{eq:scalar_shift_source}
    \varphi_{00}(t)\to\varphi_{00}(t)+2r_+c \, ,
\end{equation}
while leaving the Schwarzian and $\text{SO}(3)$ degrees of freedom unchanged.\footnote{Matching the normalizations of the $\ell = 0$ retarded correlators \eqref{eq:2pt_R} and \eqref{eq:chi_quantum} in the classical limit, using \eqref{eq:sN_1}, gives the following relation between the sources $A_{00}$ and $\varphi_{00}$ used in the classical and quantum computations, respectively: $\varphi_{00}=r_+A_{00}/\sqrt{\pi}$. Since $Y_{00}=1/\sqrt{4\pi}$, the shift \eqref{eq:scalar_shift} corresponds to $A_{00}\to A_{00}+\sqrt{4\pi}\,c$, which implies \eqref{eq:scalar_shift_source}.}

The shift symmetry \eqref{eq:scalar_shift} is also preserved by the quantum theory considered here. In the path integral with prescribed scalar boundary source, $\Phi\to\Phi+c$ is a change of integration variable with unit Jacobian. Moreover, since the transformation leaves derivatives of $\Phi$, and therefore its stress tensor, unchanged, it preserves the fixed energy and angular-momentum sectors of the microcanonical ensemble. The complete quantum source functional therefore inherits the symmetry under the corresponding constant shift \eqref{eq:scalar_shift_source}. We will use this symmetry below to constrain the allowed local counterterms and the monopole response.

\subsection{Regularized quantum response function}

Using \eqref{eq:j0_1} and \eqref{eq:j0_2}, we specialize the retarded correlator \eqref{eq:GR_time_def} to the state $|E,j=0,m_\text{\tiny BH} = 0 \rangle$ and insert a complete set of states $|E',j'=\ell,m'_\text{\tiny BH}=m \rangle$, obtaining the following expression for the Schwarzian-dressed retarded Green's function in the time domain
\begin{align}
    G_{\ell,\epsilon}^R(t) &= 2 \, \Theta(t) \, \sum_{m=-\ell}^\ell \int_{E_{0,\ell}}^{+\infty} dE' \, \rho_\ell(E') \, e^{-\frac{2\epsilon (E'-E_{0,\ell})}{E_b}} \, \left| \langle E', \ell, m | \mathcal{O}_{\ell m} | E, 0, 0 \rangle \right|^2 \sin\left( (E' - E) t\right)\nonumber\\
    \label{eq:GR_t_div}
    &= 
    2\mathcal{A}\,\Theta(t) \int_0^{+\infty} dk' \, k' \, \sinh(2 \pi k') \, \gamma_\Delta(k,k') \, e^{-\epsilon {k'}^2} \, \sin\!\left[\frac{{k'}^2 - k^2 + \ell (\ell + 1)}{2} \, E_b t \right] \, ,
\end{align}
where we used \eqref{eq:reality_lm} and \eqref{eq:CG_identity}, and defined
\begin{equation}\label{eq:A-const}
     \mathcal{A} \equiv \frac{(E_b/2)^{2 \ell + 2}}{\pi^2 \Gamma(2\ell + 1)} \,. 
\end{equation}
The parameter $\epsilon$ is a Euclidean short-time regulator for coincident insertions \cite{Mertens:2022irh}.\footnote{We use a dimensionless regulator $\epsilon$. It is related to the dimensionful short-time regulator $\epsilon$ used in \cite{Mertens:2022irh} by $\epsilon_\text{ours} = \epsilon_\text{theirs} E_b/2$.} This regulator is needed, because the $\epsilon \to 0^+$ limit of \eqref{eq:GR_t_div} is singular for every $\ell \ge 0$. The singularity comes from the short-time behaviour of the Schwarzian bilocal or, equivalently, from the UV region of the energy integral in \eqref{eq:GR_t_div}. These divergences persist even in the classical limit and are analogous to the ones we encountered in Section~\ref{subsec:retarted2pointfunction}. However, unlike in the classical computation, in the quantum case there is no unambiguous way to remove the divergences, as we discuss in Section~\ref{sec:renorm}.

To continue with our analysis, we consider the frequency domain. The Fourier transform \eqref{eq:GR_FT_def} of the regulated retarded Green's function is given by
\begin{align} \label{eq:GR_FT}
    \widehat{G}_{\ell,\epsilon}^R(\omega)
    &= \mathcal{A} \, \int_0^{+\infty} dp \, e^{-\epsilon p} \, F_\ell(p) \, \lim_{\eta \to 0^+} \int_0^{+\infty} dt \, e^{i (\omega + i \eta) t} \, \sin\!\left(\frac{p - q}{2} \, E_b t \right) \, ,
\end{align}
where we have defined the convenient shorthand notation
\begin{equation}
\label{eq:sA_p_q_Omega}
   p = {k'}^2 \, , \qquad q = k^2 - \ell (\ell + 1) \, ,
\end{equation}
\begin{equation}
\label{eq:fellp}
    F_\ell(p) = \sinh\!\left(2 \pi \sqrt{p} \right) \, \gamma_{\ell+1}\!\left(k,\sqrt{p} \right)\, . 
\end{equation}
To perform the time integral in \eqref{eq:GR_FT}, we exploit the distributional identities
\begin{equation}
\label{eq:distr_ids}
    \int_0^{+\infty} dt \, e^{i x t} = \frac{i}{x + i 0^+} = \pi \delta(x) + i \, \text{PV}\!\left(\frac{1}{x} \right) \, .
\end{equation}
This allows us to write the complete retarded Green's function as the following integral
\begin{align}
    \label{eq:GR_FT_SP}
    \widehat{G}^R_{\ell,\epsilon}(\omega) &= \frac{2 \mathcal{A}}{E_b} \, \int_0^{+\infty} dp \, e^{-\epsilon p} \, F_\ell(p) \, \frac{p-q}{(p-q)^2 - (\varpi + i 0^+)^2} \, ,
\end{align}
where
\begin{equation}
\label{eq:Omega}
    \varpi = \frac{2 \omega}{E_b}\, .
\end{equation}
On the other hand, using the second identity in \eqref{eq:distr_ids}, we can find independent formulas for the real and imaginary parts. 
The principal value term gives the real part of the retarded Green's function, while the delta function yields the imaginary part
\begin{align}
    \label{eq:Re_GR_FT}
    \text{Re}\,\widehat{G}^R_{\ell,\epsilon}(\omega) &= \frac{2 \mathcal{A}}{E_b} \, \text{PV} \int_0^{+\infty} dp \, e^{-\epsilon p} \, F_\ell(p) \, \frac{p - q}{(p-q)^2 - \varpi^2} \, ,\\
    \label{eq:Im_GR_FT_epsilon}
    \text{Im}\,\widehat{G}^R_{\ell,\epsilon}(\omega) &= \frac{\pi \mathcal{A}}{E_b} \, \left[e^{-\epsilon (q+\varpi)} \, F_\ell(q+\varpi) \, \Theta(q+\varpi) - e^{-\epsilon (q-\varpi)} \, F_\ell(q-\varpi) \, \Theta(q-\varpi) \right] \, .
\end{align}
The imaginary part has a finite $\epsilon \to 0^+$ limit,
\begin{equation}
\label{eq:Im_GR_FT}
    \text{Im}\,\widehat{G}^R_\ell(\omega) = \frac{\pi \mathcal{A}}{E_b} \, \big[F_\ell(q+\varpi) \, \Theta(q+\varpi) - F_\ell(q-\varpi) \, \Theta(q-\varpi) \big] \, ,
\end{equation}
while the real part does not, because of the divergent large-$p$ region of the integral in \eqref{eq:Re_GR_FT}.

\subsection{Classical limit}
\label{subsec:class_lim}

Before addressing the divergences of \eqref{eq:Re_GR_FT}, it is illustrative to consider the classical limit of the response function and to check that it reproduces the result in \eqref{eq:chi_result}.

Classical behavior should be recovered when the black hole energy $E$ becomes large compared to the quantum scale $E_b$, so that the quantum Schwarzian fluctuations are suppressed. In particular, to recover the classical response \eqref{eq:chi_result} we keep the semiclassical temperature \eqref{eq:beta_NE} fixed, while keeping the frequency fixed in thermal units:
\begin{equation}
\label{eq:class_lim}
    E_b \to 0 \, , \qquad k \to \infty \, , \qquad E_b k = \frac{2 \pi}{\beta} \text{ fixed} \, , \qquad \tilde{\omega} \equiv  \frac{\beta \omega}{2 \pi} \text{ fixed} \, .
\end{equation}
The multipole index $\ell$ is also held fixed as we take the limit \eqref{eq:class_lim}. We employ analytic continuation in $\ell$ in order to compute the divergent integral \eqref{eq:GR_FT_SP}. Indeed, if we remove the regulator $e^{-\epsilon p}$, the integrand in \eqref{eq:GR_FT_SP} behaves as $p^{2\ell}$ for $p\to\infty$. Thus, if we allow $\ell$ to be a real number smaller than $-1/2$, the integral converges. In fact, we consider the range
\begin{equation}
\label{eq:ell_conv}
    -1<\ell<-\frac{1}{2} \, ,
\end{equation}
where the lower bound $\ell>-1$ is imposed in order to preserve the pole structure of the integrand in a way that we make explicit below. Thus, our strategy is to evaluate \eqref{eq:GR_FT_SP} analytically in $\ell$ in the convergence interval \eqref{eq:ell_conv} and then analytically continue the resulting expression to integer values of $\ell$. We start by simplifying the form of the integral in the classical limit. First, let us make the change of integration variable
\begin{align} \label{eq:ptosdefinition}
    p=(k+s)^2\,.
\end{align}
The function $F_{\ell}(p)$ defined in \eqref{eq:fellp} then becomes
\begin{equation}
\label{eq:f_class_analytic_ell}
F_\ell\left((k+s)^2\right)=\sinh\left(2\pi(k+s)\right)\left|\Gamma(\ell+1+i(2k+s))\right|^2\left|\Gamma(\ell+1+is)\right|^2\, .
\end{equation}
Since $s$ takes values from $-k$ to infinity, it follows that $2k+s\ge k\gg 1$, and therefore, we can exploit the following asymptotic formula of the Gamma function for real numbers $\ell$ and $x$,
\begin{equation}
\label{eq:Gamma_Id}
    \left|\Gamma(\ell+1+ix)\right|^2
    \simeq 2\pi e^{-\pi|x|}|x|^{2\ell+1} \, ,
    \qquad |x|\to\infty \, .
\end{equation}
This allows us to write the integral \eqref{eq:GR_FT_SP} as 
\begin{equation}\label{GR_classical_1}
    \widehat{G}^R_\ell(\omega)
    \simeq
    \frac{4\pi\mathcal{A}}{E_b}
    \int_{-k}^{+\infty}ds\,
    e^{\pi s}
    \left|\Gamma(\ell+1+is)\right|^2\left(1-e^{-4\pi(k+s)}\right)
    \frac{s (k+s)(2k+s)^{2\ell+2}}{s^2(2k+s)^2-4k^2(\tilde{\omega}+i0^+)^2} \, ,
\end{equation}
where we have used that $\varpi=2k\tilde{\omega}$ and we have set $q=k^2$, thus neglecting $\ell(\ell+1)$ against $k^2$. We have also removed the regulator.

In order to further simplify the integral, let us observe the hierarchy $k\gg |\tilde{\omega}|$, which occurs in the classical limit as we keep $\tilde{\omega}$ fixed while taking $k\to\infty$. On the other hand, to study the large-$|s|$ behavior, we require both $|s|\gg1$, in order to use the Gamma function asymptotics \eqref{eq:Gamma_Id}, and $|s|\gg|\tilde{\omega}|$, in order to neglect the frequency-dependent term in the denominator of \eqref{GR_classical_1}. We can express these two conditions compactly as $|s|\gg1+|\tilde{\omega}|$. In this regime, the integrand in \eqref{GR_classical_1} scales as
\begin{equation}
    \text{Integrand}
    \sim
    |s|^{2\ell}(2k+s)^{2\ell}(k+s)
    \left(1-e^{-4\pi(k+s)}\right)
    \begin{cases}
        -e^{-2\pi|s|} \, , & s<0 \, ,\\
        1 \, , & s>0 \, ,
    \end{cases}
\end{equation}
where we have employed again \eqref{eq:Gamma_Id}. Since we are assuming $\ell<-1/2$, this result implies that the integrand decays faster than $1/s$ for $s\gg 1+|\tilde{\omega}|$ and is exponentially suppressed for $|s|\gg 1+|\tilde{\omega}|$ and $s<0$.

Then, we can split the integral in two regions, $|s|<s_{\rm thr}$ and $|s|>s_{\rm thr}$ for a certain threshold value $s_{\rm thr}$. The previous result implies that the contribution from the region $|s|>s_{\rm thr}$ to \eqref{GR_classical_1} vanishes as we take $s_{\rm thr}\gg1+|\tilde{\omega}|$. The crucial observation is that this suppression holds independently of the size of $s_{\rm thr}$ relative to $k$: at fixed $E_b k$, the $k$-dependent factors in \eqref{GR_classical_1} at large $|s|$, including the overall prefactor $\mathcal{A}/E_b$, combine as
\begin{equation}
    \frac{\mathcal{A}}{E_b}(2k+s)^{2\ell}(k+s)
    \left(1-e^{-4\pi(k+s)}\right)
    \propto
    \left(2+\frac{s}{k}\right)^{2\ell}
    \left(1+\frac{s}{k}\right)
    \left(1-e^{-4\pi(k+s)}\right) \, ,
\end{equation}
which is bounded throughout the integration domain $s\ge-k$ when $\ell<-1/2$. We can then choose a threshold satisfying the hierarchy $1+|\tilde{\omega}|\ll s_{\rm thr}\ll k$ and take $k\to\infty$ and $s_{\rm thr}\to\infty$ while maintaining the hierarchy. Hence, the integral is confined to the region of $|s|\ll k$ and \eqref{GR_classical_1} is reduced to a simpler integral
\begin{equation}
    \widehat{G}^R_\ell(\omega)
    \simeq
    \frac{2\pi\mathcal{A}}{E_b}(2k)^{2\ell+1}
    \int_{-\infty}^{+\infty}ds\,
    e^{\pi s}
    \left|\Gamma(\ell+1+is)\right|^2
    \frac{s}{s^2-(\tilde{\omega}+i0^+)^2} \, .
\end{equation}
Let us now observe that, since $e^{\pi s}$ is multiplied by an odd function and the integration range is symmetric, we can do the replacement $e^{\pi s}\to \sinh\left(\pi s\right)$. Then, we can make use of the following identity
\begin{equation}
    \sinh(\pi s)\left|\Gamma(\ell+1+is)\right|^2=
\frac{\pi}{2i\cos(\pi\ell)}\left[\frac{\Gamma(\ell+1+is)}{\Gamma(-\ell+is)}-\frac{\Gamma(\ell+1-is)}{\Gamma(-\ell-is)}\right]\, ,
\end{equation}
which can be straightforwardly derived from Euler's reflection formula for the Gamma function. The two ratios of Gamma functions give opposite contributions to the integral, as follows by changing $s\to-s$. We thus reduce the integral to
\begin{equation}
    \widehat{G}^R_\ell(\omega)
    \simeq
    -\frac{2\pi^2\mathcal{A}(2k)^{2\ell+1}}{i E_b \cos(\pi\ell)}
    \int_{-\infty}^{+\infty}ds\,
    \frac{\Gamma(\ell+1-is)}{\Gamma(-\ell-is)}
    \frac{s}{s^2-(\tilde{\omega}+i0^+)^2} \, .
\end{equation}
We can then evaluate this integral via the residue theorem, by closing the contour with an arc of radius $R\to\infty$ in the upper half plane. For $s=R e^{ i \theta}$ with $\theta\in[0, \pi]$, the ratio of Gamma functions scales as $R^{2\ell+1}$, and the integrand scales as $R^{2\ell}$. Multiplying this by the length of the arc, we find that the integral over the arc scales as $R^{2\ell+1}$, which vanishes for $R\to\infty$ since we are taking $\ell<-1/2$. Thus, the integral is reduced to a sum of residues in the upper half plane. The Gamma function $\Gamma(\ell+1-is)$ has poles at $s=-i(n+\ell+1)$ for $n=0,\, 1,\, 2,\, \ldots$. For $-1<\ell<-1/2$, all of these are in the lower half plane and they do not contribute to the integral. Hence, the residue sum contains a single pole: the one at $s=\tilde{\omega}+i 0^{+}$ coming from the rational part of the integrand. The evaluation of the residue is straightforward, and we get
\begin{equation}
    \widehat{G}^R_\ell(\omega)
    \simeq
    -\frac{2\pi^3\mathcal{A}(2k)^{2\ell+1}}{E_b \cos(\pi\ell)}
    \frac{\Gamma(\ell+1-i\tilde\omega)}{\Gamma(-\ell-i\tilde\omega)}\, .
\end{equation}
Inserting the value of the constant $\mathcal{A}$ defined in \eqref{eq:A-const} and using that $E_bk=2\pi/\beta$, we finally get
\begin{equation}\label{eq:GR_class_analytic_ell_result}
    \widehat{G}^R_\ell(\omega)
    \stackrel{\text{cl}}{\longrightarrow}
    \frac{\pi}{\Gamma(2\ell + 1)\cos[\pi(\ell+1)]}\left(\frac{2\pi}{\beta}\right)^{2\ell+1}
    \frac{\Gamma(\ell+1-i\tilde\omega)}{\Gamma(-\ell-i\tilde\omega)}\, .
\end{equation}
Having obtained \eqref{eq:GR_class_analytic_ell_result} in the convergence range $-1<\ell<-1/2$, we now analytically continue this expression to arbitrary values of $\ell$. Multiplying by the normalization \eqref{eq:sN_1} and using standard Gamma-function identities, we find that in the classical limit the quantum response function \eqref{eq:chi_quantum} approaches
\begin{equation}
    \hat{\chi}_\ell(\omega)
    \stackrel{\text{cl}}{\longrightarrow}
    \frac{1}{2}
    \left(\frac{\pi r_+}{\beta}\right)^{2\ell+1}
    \frac{
        \Gamma\!\left(-\ell-\frac{1}{2}\right)
        \Gamma(\ell+1-i\tilde{\omega})
    }{
        \Gamma\!\left(\ell+\frac{1}{2}\right)
        \Gamma(-\ell-i\tilde{\omega})
    } \, .
\end{equation}
This is precisely the classical response function \eqref{eq:chi_result}.
The analytic continuation procedure thus reconstructs the full analytic-in-$\ell$ classical response. The corresponding matching of finite local terms at integer $\ell$ is discussed in Appendix~\ref{app:class_lim}.

\subsection{Renormalization}
\label{sec:renorm}
Let us now address the evaluation of \eqref{eq:GR_FT_SP} --- or more precisely, its real part \eqref{eq:Re_GR_FT} --- in the quantum regime. 
A priori, it is possible to perform the same strategy as we just did in the classical limit: we evaluate the integral in the convergence range $-1<\ell<-1/2$ and then we analytically continue the result to integer $\ell$. However, there are two problems with this approach. 
First, there is a practical obstruction. This method requires that we are able to obtain a closed-form analytic expression in $\ell$, which is very challenging due to the complicated form of the integrand. Second, there is a conceptual concern. Analytic continuation gives us an unambiguous answer for the response function: provided we can find the result in the range $-1<\ell<-1/2$, its analytic extension is unique. In the classical limit, this correctly reproduces the classical response function, and this is reasonable, because the classical computation in Section \ref{sec:CLN} also relies on analytic continuation, and is scheme independent. 
However, in the quantum case, this apparent absence of ambiguities might be deceiving. In fact, in general the result can depend on the choice of a renormalization scheme, implemented through the choice of counterterms. In this sense, analytic continuation in $\ell$ would be providing us with a particular renormalization scheme, but without further analysis (\textit{i.e.} by matching the result to a scattering and worldline effective field theory approach) it may be unjustified to attribute it a preferred physical significance.

Therefore, in this section we address the renormalization of \eqref{eq:Re_GR_FT} by examining the divergences for $\epsilon\to 0$ and removing them with appropriate counterterms. 

From the start, we assume that $\ell$ takes integer values. Starting from the bare expression for the real part of the retarded Green's function, \eqref{eq:Re_GR_FT}, we first focus our attention on the function $F_\ell(p)$ defined in \eqref{eq:fellp} entering the integral, which is responsible for the UV divergences. For integer $\ell \in \mathbb{N} = \{0,1,2,\ldots\}$, the Schwarzian matrix element can be written as
\begin{align}\notag
    \gamma_{\ell+1}(k,k') &= \big|\Gamma\big(\ell + 1 + i (k+k') \big) \big|^2 \, \big|\Gamma\big(\ell + 1 + i (k-k') \big) \big|^2\\
    &= \frac{\pi^2 \, ({k'}^2 - k^2)}{\sinh^2(\pi k') - \sinh^2(\pi k)} \, \prod_{n=1}^\ell \left[n^2 + (k'+k)^2 \right] \, \left[n^2 + (k'-k)^2 \right] \, ,
    \label{eq:matrix_element_polynomial}
\end{align}
where in the last step we exploited the identity
\begin{equation}
    |\Gamma(\ell + 1 + i y)|^2 = \frac{\pi y}{\sinh(\pi y)} \, \prod_{n=1}^\ell (n^2 + y^2) \, ,
\end{equation}
for integers $\ell \geq 0$. For $\ell=0$, the product is unity. As a consequence, the function $F_\ell(p)$ given in \eqref{eq:fellp} can be written as
\begin{equation} \label{eq:fwithQandsinh}
    F_\ell(p) = \pi^2 \, Q_\ell(p) \, \frac{\sinh\!\left(2 \pi \sqrt{p} \right)}{\sinh^2\!\left(\pi \sqrt{p} \right) - \sinh^2(\pi k)} \, ,
\end{equation}
where we have defined the real polynomial $Q_\ell$ of degree $(2 \ell + 1)$ given by
\begin{align}
    \label{eq:Qell}
    Q_\ell(p) &= (p - k^2) \, \prod_{n=1}^\ell \left[p^2 + 2 (n^2 - k^2) p + (n^2 + k^2)^2 \right]\,.
\end{align}
Since $Q_\ell(p)$ is a polynomial in $p$, it is convenient to express it as
\begin{align}
    \label{eq:Qell_app_2}
    Q_\ell(p) &= \sum_{i=0}^{2 \ell + 1} b_i^{} \, p^i \, ,
\end{align}
where $b_{i}$ are coefficients that depend on $\ell$ and $k$. To understand the structure of the divergences that appear in \eqref{eq:Re_GR_FT}, let us consider how \eqref{eq:fwithQandsinh} behaves as $p \to +\infty$. The ratio of hyperbolic sine functions gives
\begin{equation}
    \frac{\sinh\!\left(2 \pi \sqrt{p} \right)}{\sinh^2\!\left(\pi \sqrt{p} \right) - \sinh^2(\pi k)} = 2 + O\!\left(e^{-2 \pi \sqrt{p}} \right) \, .
\end{equation}
Hence, we can split $F_\ell(p)$ as
\begin{equation}
\label{eq:sinh_gamma}
    F_\ell(p) = 2 \pi^2 Q_\ell(p) + R_\ell(p) \, ,
\end{equation}
where the remainder
\begin{equation}
\label{eq:Rell}
    R_\ell(p) = \pi^2 \, Q_\ell(p) \, \left[\frac{\sinh\!\left(2 \pi \sqrt{p} \right)}{\sinh^2\!\left(\pi \sqrt{p} \right) - \sinh^2(\pi k)} -2 \right]\,,
\end{equation}
is exponentially suppressed as $p \to +\infty$. In this form, all the UV divergences arise only from the term proportional to $Q_\ell$ in \eqref{eq:sinh_gamma}. Although the energy integral in \eqref{eq:Re_GR_FT} extends formally to arbitrarily high intermediate energies, the matrix elements \eqref{eq:matrix_elements} hold only for transitions with $|E'-E|\ll 1/r_+$, which do not resolve the mouth of the throat.
As in an effective theory, the resulting UV divergences are removed by local counterterms, as we show below, and their finite parts are in general matching data of the four-dimensional theory.

Applying the decomposition \eqref{eq:sinh_gamma} to the real part of the Green's function in \eqref{eq:Re_GR_FT}, we obtain
\begin{align} \label{eq:ReGRform2}
    \text{Re}\,\widehat{G}^R_{\ell,\epsilon}(\omega) =~& \frac{ \mathcal{A}}{E_b} \sum_{\sigma=\pm1}\left[ \, \text{PV} \int_0^{+\infty} dp \, e^{-\epsilon p} \, \left(2 \pi^2 \sum_{i=0}^{2 \ell + 1} b_i p^i + R_\ell(p)\right)  \, \frac{1}{p-q-\sigma \varpi}\, \right] \,,
\end{align}
where we have used
\begin{equation}
    \frac{p-q}{(p-q)^2 - \varpi^2} = \frac{1}{2} \, \sum_{\sigma=\pm1} \frac{1}{p-q-\sigma \varpi} \,,
\end{equation}
and the expression of $Q_\ell(p)$ given in \eqref{eq:Qell_app_2}.
We first evaluate the integral of the $Q_\ell(p)$ term; the convergent integral of $R_\ell(p)$ is evaluated at the end of this section. To do so, we note that
\begin{align}
\label{eq:basic_integral_app}
    \text{PV} \int_0^{+\infty} dp \, e^{-\epsilon p} \, \frac{p^i}{p-a} = \begin{cases}
          -\gamma_E - \log(\epsilon |a|) & i=0\,,
          \\
          a^{i}\left(\sum_{n=1}^{i} \frac{\Gamma(n)}{(a \epsilon)^{n}} - \gamma_E - \log(\epsilon |a|)\right)  & i>0\,,
       \end{cases}
\end{align}
where $a \neq 0$ and we have perturbatively expanded for small $\epsilon$. Thus, we find
\begin{align} \label{eq:Re_GR_app}
    \text{Re}\,\widehat{G}^R_{\ell,\epsilon}(\omega) =~& \frac{ \mathcal{A}}{E_b} \sum_{\sigma=\pm1}\, \left[ - 2 \pi^2Q_\ell(q + \sigma \varpi)(\gamma_E + \log (\epsilon |q + \sigma \varpi|)) + 2 \pi^2\sum_{i=1}^{2\ell+1}\sum_{n=1}^{i} \frac{b_{i}\Gamma(n)\epsilon^{-n}}{(q+\sigma \varpi)^{n-i}}  \nonumber \right. \\ & \left. + \text{PV} \int_0^{+\infty} dp \,e^{-\epsilon p} R_\ell(p) \, \frac{1}{p-q-\sigma \varpi} \right]\, .
\end{align}
The form of the divergences changes depending on the value of the multipole index. For $\ell\geq 1$, the regulated real part contains power-law divergences $\epsilon^{-(2 \ell + 1)},\ldots,\epsilon^{-1}$ as well as logarithmic divergences of the form $\log(\epsilon)$. On the other hand, the case $\ell=0$ is special in that the logarithmic divergence cancels and only the $\epsilon^{-1}$ divergence remains.
In fact, the coefficient of $\log(\epsilon)$ for $\ell=0$ is proportional to $Q_0(q + \varpi)+Q_0(q - \varpi)=2(q-k^2)$, which vanishes because $q=k^2$ for $\ell=0$ --- see \eqref{eq:sA_p_q_Omega}.

For $\ell\ge 1$, the presence of the logarithmic divergence implies RG running of the response function. The physical significance of this term can be seen even before subtracting the divergences, as the coefficient of $\log(\epsilon)$ is independent of the choice of regulator --- for instance, it remains invariant if we perform a rescaling $\epsilon\to A \epsilon$, while the finite terms are ambiguous. 

To see this explicitly, we renormalize the result. The divergences in \eqref{eq:Re_GR_app} allow us to see directly the counterterms we should include in the Schwarzian action in order to get a finite answer. We have 
\begin{align}
    \label{eq:Ict_t}
    I_\text{ct}
    = \frac{1}{2 (2 \ell + 1)} \, \sum_{m=-\ell}^\ell \int dt \, \varphi_{\ell m}^\ast(t) \, K^{\text{ct}}_{\ell} \, \varphi_{\ell m}(t) \, ,
\end{align}
where\footnote{We are choosing to absorb the Euler--Mascheroni constant $\gamma_E$ in the counterterms when we renormalize the retarded Green's function. This changes only the finite part and leaves the beta function unchanged.}
\begin{align}\notag
    K^{\text{ct}}_{\ell}=& \frac{2\pi^2\mathcal{A}}{E_b} \sum_{\sigma=\pm1} \, \left[Q_\ell(\boldsymbol{q} + \sigma \boldsymbol{\varpi}) \, \left(\gamma_E + \log(\epsilon \mu) \right) - \sum_{i=1}^{2\ell+1} \sum_{n=1}^{i} b_{i}(\boldsymbol{k}^2) \, (\boldsymbol{q}+\sigma \boldsymbol{\varpi})^{i-n}\frac{\Gamma(n)}{\epsilon^{n}}  \right]\\
    &+ P_\ell(-\partial_{t}^2, \boldsymbol{k}^2) \, .
    \label{eq:Kct_MS}
\end{align}
Here, the boldface symbols denote operators, so that the counterterms \eqref{eq:Kct_MS} are independent of the chosen microcanonical energy state. Explicitly,
\begin{equation}
    \boldsymbol{k}^2=\frac{2H}{E_b}\, ,\qquad \boldsymbol{q}=\boldsymbol{k}^2-\ell(\ell+1)\, ,\qquad \boldsymbol{\varpi}=\frac{2i}{E_b}\partial_t\, ,
\end{equation}
where $H$ is the unperturbed Schwarzian Hamiltonian and the time derivatives act on the external scalar source. All occurrences of $k^2$ in $Q_\ell$ and in the coefficients $b_i$ are understood to be promoted to $\boldsymbol{k}^2$. Crucially, the counterterms \eqref{eq:Kct_MS} are polynomial in the Schwarzian Hamiltonian $H \propto \boldsymbol{k}^2$ and in time derivatives $i \partial_t$ of the source, and can therefore be represented by local operators in the Schwarzian effective theory.

In addition, $\mu$ is a dimensionless renormalization scale, while $P_\ell(-\partial_{t}^2, \boldsymbol{k}^2)$ represents finite local counterterms. We work at finite order in the local operator expansion in powers of the Schwarzian Hamiltonian and time derivatives of the source, so $P_\ell$ is a polynomial in both arguments. These parametrize the freedom to choose a renormalization scheme, subject to the constraints imposed by symmetries and classical matching. In the time domain, these local counterterms contribute to the response through contact terms supported at $t=0$.

When we add the counterterms to \eqref{eq:Re_GR_app} and take $\epsilon\to0$, the power-law divergences cancel and the logarithmic divergence is traded for the scale $\mu$. In the integral of $R_\ell(p)$, which converges since $R_\ell(p)$ decays exponentially, the limit simply removes the factor $e^{-\epsilon p}$. The real part of the renormalized response is then
\begin{align} \label{eq:Re_GR_ren}
    \text{Re}\,\widehat{G}^{R,\text{ren}}_\ell(\omega) &= \frac{ 2 \pi^2\mathcal{A}}{E_b} \sum_{\sigma = \pm 1} \left[ Q_\ell(q + \sigma \varpi) \log \frac{\mu}{|q + \sigma \varpi|} \nonumber
    \right. \\& \left. \qquad \qquad \qquad \quad + \frac{1}{2\pi^2}\text{PV} \int_0^{+\infty} dp \frac{R_\ell(p)}{p-q-\sigma \varpi} \right]+ P_\ell(\omega^2,k^2) \, .
\end{align}
The polynomial $P_\ell(\omega^2,k^2)$ reflects the freedom in choosing a renormalization scheme, which in general makes the result ambiguous. An exception occurs in the static $\ell = 0$ sector, where the renormalization freedom is constrained by the scalar shift symmetry described in \eqref{eq:scalar_shift} and \eqref{eq:scalar_shift_source}, which we discuss in detail in Section~\ref{sec:QSLN}. 
For $\ell\geq1$, the conservative response remains subject to renormalization ambiguity but we can determine its running with the scale $\mu$.
By contrast, we recall that the imaginary part \eqref{eq:Im_GR_FT} is finite and scheme independent.

The integral of $R_\ell$ in \eqref{eq:Re_GR_ren} can be done in closed form, as we show in Appendix~\ref{app:nonpol}. Its logarithm cancels the $\log|q+\sigma\varpi|$ of the first term, and the result gives the real and imaginary parts of the response together,
\begin{equation}
\label{eq:GR_closed}
    \widehat{G}^{R,\text{ren}}_\ell(\omega) = \frac{2\pi^2\mathcal{A}}{E_b} \sum_{\sigma=\pm1} Q_\ell(q+\sigma\varpi) \Big[ \log\mu - \Psi_k(q+\sigma\varpi) \Big] + \mathrm{Pol} \, ,
\end{equation}
where
\begin{equation}
\label{eq:Psi_k}
    \Psi_k(a) = \psi\big(\sqrt{-a}+ik\big) + \psi\big(\sqrt{-a}-ik\big) - \frac{\sqrt{-a}}{a-k^2}
\end{equation}
and $\psi(z)=\Gamma'(z)/\Gamma(z)$ is the digamma function. Here $\mathrm{Pol}$ is a real polynomial in $\omega^2$ and $k^2$ that includes $P_\ell$, and the square roots are on their principal branch with $\varpi\to\varpi+i0^+$, that is, $\sqrt{-q-\varpi}=-i\sqrt{q+\varpi}$ when $q+\varpi>0$ and $\sqrt{-q+\varpi}=+i\sqrt{q-\varpi}$ when $q-\varpi>0$. The imaginary part of \eqref{eq:GR_closed} is \eqref{eq:Im_GR_FT}.

Two properties of \eqref{eq:GR_closed} will be important. First, the response has no logarithmic singularity at the thresholds $q+\sigma\varpi=0$, only square-root branch points. Second, it shows how the logarithm of the renormalization scale combines with the physical scales. The coefficient of $\log\mu$ determines the RG running at fixed finite local terms, while the digamma functions determine the accompanying frequency and energy dependence. Introducing the dimensionful renormalization scale $\bar{\mu}=E_b\mu/2$, we can display the logarithmic scale dependence of the conservative response as
\begin{align}
\label{eq:Psi_scales}
    \text{Re}\,\widehat{G}^{R,\text{ren}}_\ell(\omega)
    &=
    \frac{2\pi^2\mathcal{A}}{E_b}
    \sum_{\sigma=\pm1}Q_\ell(q+\sigma\varpi)
    \nonumber\\
    &\qquad \times
    \left[
        \log\frac{\bar{\mu}}
        {\max\left\{|\omega|,\sqrt{2E_bE},E_{0,\ell}\right\}}
        + O(1)
    \right]
    + \mathrm{Pol} \, .
\end{align}
This estimate holds for $\ell\ge1$, provided $|\omega|$ is not parametrically close to $E_{0,\ell}$. The terms of order one can depend non-polynomially on the scale ratios and are not included in $\mathrm{Pol}$.

When the frequency dominates, the logarithm compares $\bar{\mu}$ with $|\omega|$. When $\sqrt{2E_bE}$ dominates, the relevant scale is set by the semiclassical temperature, $2\pi/\beta=\sqrt{2E_bE}$, rather than $E$ itself.\footnote{The reason is that the classical response is a function of $\beta\omega$, and not of $\omega/E$. More precisely, the response is a sum over transitions, and these only know about energy differences $E'-E$. In the semiclassical regime, the density of states at nearby energies satisfies $\rho(E')/\rho(E)\simeq e^{\beta(E'-E)}$, so $1/\beta$ is the characteristic energy scale.} When the rotational energy dominates, the scale entering the logarithm is instead set by $E_{0,\ell}$. Since $E_{0,\ell}$ depends neither on $E$ nor on $\omega$, the logarithm is then a constant.
Even in this rotational-energy-dominated regime, the exact response retains its non-polynomial dependence on energy and frequency. As explained above, this discussion assumes $E\ll S_0/r_+$ and $r_+|\omega|\ll1$.

\section{Quantum Love numbers}
\label{sec:QSLN}

\subsection{Static Love numbers}
\label{subsec:StLN}
Using the renormalized Green's function found in \eqref{eq:Re_GR_ren} and setting $\omega\to 0$, we find the following expression for the static quantum Love numbers, defined by \eqref{eq:QLN_def}:
\begin{align} \label{eq:SLN_ren}
    \lambda_\ell &= \frac{(2 \ell + 1) \, \ell! \, (\ell + 1)!}{\pi \, \Gamma(2 \ell + 2) \, \Gamma(2 \ell + 3)} \, (r_+ E_b)^{2 \ell + 1} \, \left(Q_\ell(q) \, \log \frac{\mu}{|q|} + \mathcal{I}_\ell \right) + \frac{\Gamma(\ell+1)r_+^{2\ell+1}}
{4\sqrt{\pi}\,\Gamma\!\left(\ell+\frac32\right)}
\, P_\ell(0,k^2) \, ,
\end{align}
where
\begin{equation}
\label{eq:sI}
    \mathcal{I}_\ell = \text{PV} \int_0^{+\infty} dk' \, k' \, \frac{Q_\ell({k'}^2)}{{k'}^2-q} \, \left(\frac{\sinh(2 \pi k')}{\sinh^2(\pi k') - \sinh^2(\pi k)} - 2 \right) \, .
\end{equation}
The role of the rotational energy in producing nonzero Love numbers can be seen by formally dropping the offset $\ell(\ell+1)$ in $q$, which is where $E_{0,\ell}$ enters the energy differences, while keeping $F_\ell(p)$ unchanged. Then $q=k^2$ and $Q_\ell(q)=0$, so the logarithmic term vanishes. By \eqref{eq:GR_closed}, the same factor $Q_\ell(q)$ multiplies all the non-polynomial dependence of the static response, and what remains is a polynomial in $k^2$, which can be absorbed into the finite local terms. The non-polynomial static response and the logarithmic running therefore require the rotational offset $\ell(\ell+1)$. For $\ell=0$ there is no offset. Hence the $\ell = 0$ and $\ell \ge 1$ cases differ, as we now describe.

\subsubsection*{Case: $\ell =0$}
For $\ell=0$, the function $Q_0(q)=0$, so the logarithmic term in \eqref{eq:SLN_ren} vanishes and the static Love number $\lambda_0$ does not run with $\mu$. In this case, we can evaluate $\lambda_0$ more explicitly:
\begin{align}
    \lambda_0
    &= \frac{r_+ E_b}{2 \pi} \, \text{PV} \int_0^{+\infty} dk' \, k' \,
    \left(
        \frac{\sinh(2 \pi k')}
        {\sinh^2(\pi k')-\sinh^2(\pi k)}
        -2
    \right)
    +\frac{r_+}{2\pi} P_0(0,k^2)
    \nonumber\\
    \label{eq:lambda_0}
    &= \frac{r_+ E_b}{2 \pi}
    \left(\frac{1}{6}-k^2\right)
    +\frac{r_+}{2\pi}P_0(0,k^2) \, .
\end{align}
We can see that the static Love number depends on $P_0(0,k^2)$; however, the shift symmetry condition \eqref{eq:scalar_shift_source} constrains the response. Since the renormalized source functional is invariant under $\varphi_{00}(t)\to\varphi_{00}(t)+a$, with $a$ constant, the expectation value of $\mathcal{O}_{00}$ is unchanged under this transformation. In linear response this implies
\begin{equation}
    a\int dt'\,G^{R,\mathrm{ren}}_0(t-t') = 0 \, ,
\end{equation}
and therefore, by time-translation invariance, $\widehat{G}^{R,\mathrm{ren}}_0(0)=0$. That is
\begin{equation}
\label{eq:shift_Ward_GR}
    \lambda_0=0 \, .
\end{equation}
Combining \eqref{eq:shift_Ward_GR} with \eqref{eq:lambda_0} fixes the counterterm $P_0(0,k^2)$ to
\begin{equation}
\label{eq:P0_shift}
    P_0(0,k^2) = E_b\left(k^2-\frac{1}{6}\right) \, .
\end{equation}
Any additional finite $\ell = 0$ counterterm must satisfy $\delta P_0(0,k^2)=0$, whereas the coefficients of $\omega^2,\omega^4,\ldots$ remain unconstrained by the shift symmetry. For $\ell\geq1$, the shift symmetry imposes no corresponding constraint. 

\subsubsection*{Case: $\ell\geq1$}

For $\ell\geq1$, $Q_\ell(q)$ is nonzero, and the logarithm of the renormalization scale $\mu$ in \eqref{eq:SLN_ren} induces RG running of the static Love number $\lambda_\ell$. With the local term $P_\ell$ held fixed, changing the renormalization scale from $\mu_0$ to $\mu$ gives
\begin{equation}
\label{eq:lambda_ell_RG}
    \lambda_\ell(\mu) = \lambda_\ell(\mu_0) + \text{B}_\ell \log\frac{\mu}{\mu_0} ,
\end{equation}
where the finite reference value $\lambda_\ell(\mu_0)$ contains matching coefficients that are not determined by the near-horizon theory and must be fixed by a renormalization condition. While the finite value of $\lambda_\ell(\mu)$ is scheme dependent, the beta function $\text{B}_\ell$ is scheme independent and is given by
\begin{align}
    \label{eq:B_ell_1}
    \text{B}_\ell
    &=
    \mu \, \frac{d\lambda_\ell}{d\mu}
    = -\frac{\ell \, (2 \ell + 1) \, \Gamma(\ell + 2)^2}{\pi \, \Gamma(2 \ell + 2) \, \Gamma(2 \ell + 3)} \, (r_+ E_b)^{2 \ell + 1} \, \prod_{n=1}^\ell \left[\left(n^2 - \ell (\ell + 1) \right)^2 + (2 k n)^2 \right] \,, \quad \ell \geq 1\,,
\end{align}
where we have used \eqref{eq:Qell}. For example,
\begin{align}
\label{eq:B1}
    \text{B}_1 &= - \frac{(r_+ E_b)^3}{12 \pi} \left(\frac{8 E}{E_b} + 1 \right) \, ,
    \nonumber \\
    \text{B}_2 &= -\frac{(r_+ E_b)^5}{60 \pi} \, \left(\frac{8 E}{E_b} + 1 \right) \, \left(\frac{8 E}{E_b} + 25 \right) \, ,
    \nonumber \\
    \text{B}_3 &= -\frac{3 \, (r_+ E_b)^7}{175 \pi} \, \left(\frac{8 E}{E_b} + 1 \right) \, \left(\frac{8 E}{E_b} + 121 \right) \, \left(\frac{E}{E_b} + 2 \right) \, ,
\end{align}
while for $\ell=0$ there is no running, $\text{B}_0=0$. For $\ell \ge 1$, the beta function $\text{B}_\ell$ is always negative and therefore, with $P_\ell$ held fixed, $\lambda_\ell(\mu)$ decreases as the renormalization scale is increased.

The classical matching, as we describe in Appendix~\ref{app:class_lim}, fixes the asymptotic behavior of the finite local counterterms $P_{\ell}(0,k^2)$ required to recover the classical static response and therefore gives $\lambda_\ell \to 0$ for any $\ell \ge 0$. We remark that this does not remove the finite renormalization ambiguity of the $\ell \ge 1$ quantum static Love numbers at fixed $E_b$ and $k$.

Equation \eqref{eq:lambda_ell_RG} describes the running with $\mu$ when the finite local terms are held fixed. These terms must ultimately be fixed by matching the near-horizon response to the full four-dimensional theory. The renormalized coefficients of these finite local terms can depend on $\mu$, compensating the running of the near-horizon contribution so that physical observables are independent of the arbitrary renormalization scale. This is possible because $\text{B}_\ell$ is a polynomial in the energy. What the near-horizon theory fixes is the dependence on the energy up to local polynomial terms. Setting $\omega=0$ in \eqref{eq:GR_closed}, we find
\begin{align}
\label{eq:lambda_ell_closed}
    \lambda_\ell &= \text{B}_\ell \left[ \log\mu - L_\ell(E) \right] + \mathrm{Pol}_\ell(E) \, , \nonumber\\
    L_\ell(E) &= \text{Re}\,\Psi_k(q) = \text{Re} \left[ \psi\!\left( \sqrt{-q} + i k \right) + \psi\!\left( \sqrt{-q} - i k \right) \right] + \frac{\text{Re}\sqrt{-q}}{\ell(\ell+1)} \, ,
\end{align}
where $\mathrm{Pol}_\ell(E)$ is a polynomial that includes the local terms. The static version of the logarithmic scale estimate \eqref{eq:Psi_scales} is
\begin{equation}
\label{eq:lambda_ell_scales}
    \lambda_\ell = \text{B}_\ell \left[ \log\frac{\bar{\mu}}{\max\left\{\sqrt{2E_bE},E_{0,\ell}\right\}} + O(1) \right] + \mathrm{Pol}_\ell(E) \, .
\end{equation}
Here $\bar{\mu}=E_b\mu/2$, as in \eqref{eq:Psi_scales}, and the terms of order one can depend non-polynomially on the energy and are not included in $\mathrm{Pol}_\ell(E)$.

When $\sqrt{2E_bE}\gg E_{0,\ell}$, the scale entering the logarithm is set by the semiclassical temperature \eqref{eq:beta_NE}. More precisely, for fixed $\ell\ge1$ and $k\gg\ell(\ell+1)$, Appendix~\ref{app:nonpol} gives $L_\ell(E)=\log(2k)-\gamma_E+O(k^{-2})$. Using $2k=4\pi/(\beta E_b)$, we obtain
\begin{equation}
\label{eq:lambda_ell_thermal}
    \lambda_\ell = \text{B}_\ell \left[ \log\frac{\beta\bar{\mu}}{2\pi} + O(k^{-2}) \right] + \mathrm{Pol}_\ell(E) \, ,
\end{equation}
where the polynomial term $\gamma_E\text{B}_\ell$ has been absorbed into $\mathrm{Pol}_\ell(E)$. Thus, the physical scale entering the logarithm is set by $1/\beta$, and the coefficient of the explicit $\log E$ term is $-\text{B}_\ell/2$. However, this semiclassical limit must still be taken within the near-extremal regime, $E\ll S_0/r_+$, or equivalently $r_+/\beta\ll1$. Our calculation therefore does not extend to arbitrarily high energies or temperatures.

When the rotational energy dominates instead, the scale entering the logarithm is set by $E_{0,\ell}$. In particular, the exact function $L_\ell(E)$ tends to the constant $L_\ell(0)=2\psi\big(\sqrt{\ell(\ell+1)}\big)+1/\sqrt{\ell(\ell+1)}$ as $E\to0$, so no logarithm of $E$ develops in this limit. Since $\text{B}_\ell L_\ell$ is not a polynomial in $E$, no choice of the finite local counterterms can make $\lambda_\ell$ vanish identically as a function of $E$.

\subsection{Dissipative Love numbers}
\label{sec:QDLN}
\begin{figure}[t!]
    \centering
    \includegraphics[width=0.475\textwidth]{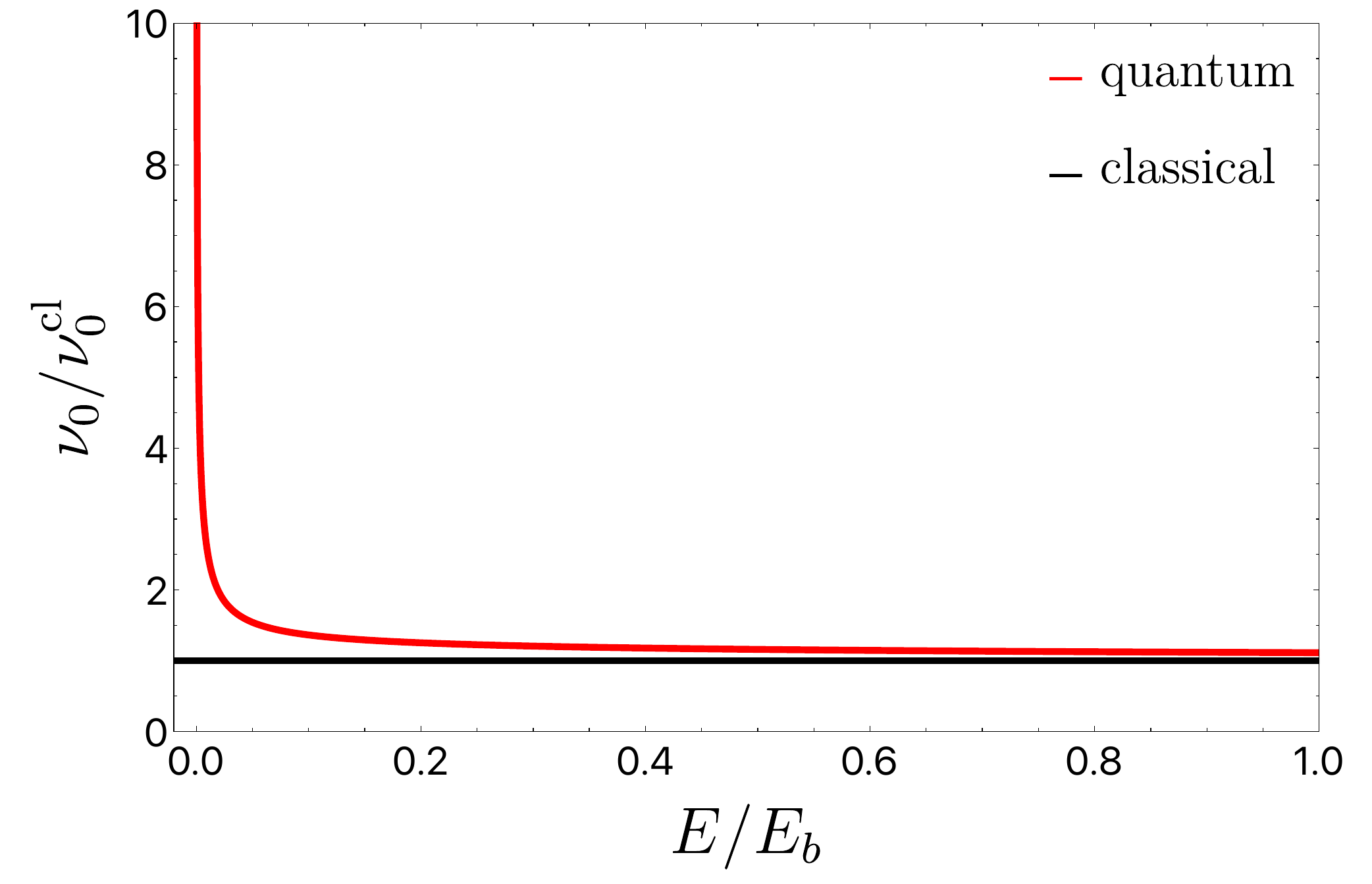}\qquad\includegraphics[width=0.475\textwidth]{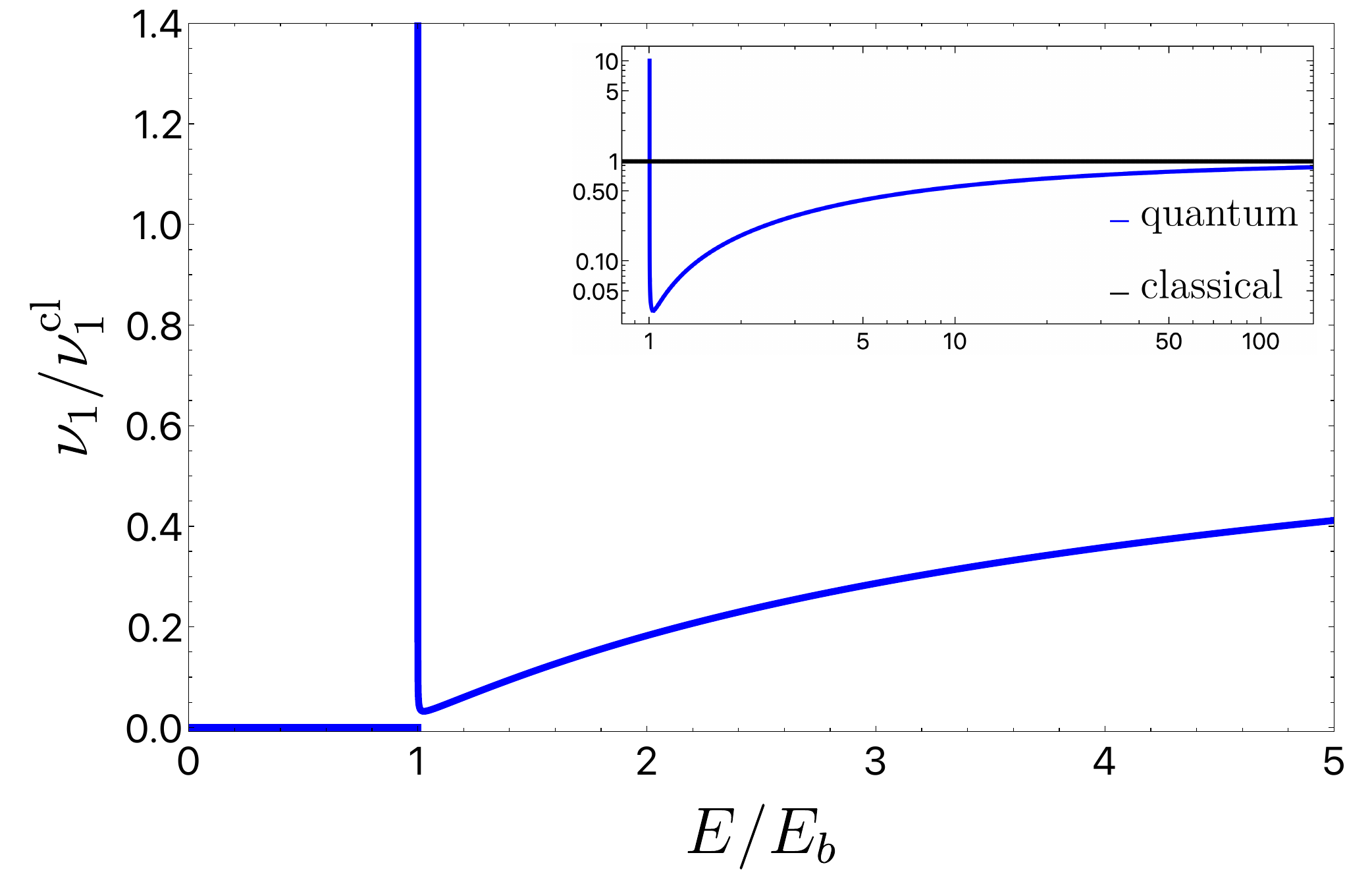}
    \caption{\small We show the quantum dissipative scalar Love numbers \eqref{eq:nu_ell} for $\ell = 0$ (left) and $\ell = 1$ (right), normalized by their classical values  \eqref{eq:nu_cl}, as a function of $E/E_b$. The quantum results \eqref{eq:nu0} (red) and \eqref{eq:nu1} (blue) approach the classical value $\nu_\ell/\nu^\text{cl}_\ell=1$ (black) for $E/E_b \gg 1$. In the right panel, this is shown in the log-log inset. For $\ell = 0$ the classical value is approached from above, while for $\ell = 1$ it is approached from below, corresponding respectively to an enhancement and a suppression of the low-frequency absorption cross-section (cf.~\eqref{eq:nu_ell_sigma_abs}). For $\ell = 1$, the dissipative Love number vanishes in the energy window $0<E/E_b<1$, while no analogous gap is present for $\ell = 0$. Both curves diverge as the corresponding threshold $E - E_{0,\ell} \to 0^+$ as described in \eqref{eq:nu_ratio_small_q}.}
    \label{fig:nu0_nu1}
\end{figure}

To find the dissipative Love numbers, we perturbatively expand the imaginary part of the retarded Green's function \eqref{eq:Im_GR_FT} for small $\omega$.
This first requires us to understand how the function behaves for different energy levels. For $q < 0$, both step functions in \eqref{eq:Im_GR_FT} vanish for sufficiently small $\omega$, and hence the imaginary response also vanishes as $\omega \to 0$. On the other hand, for $q>0$, the step functions in \eqref{eq:Im_GR_FT} equal one for sufficiently small $\omega$ and we find
\begin{equation}
    \text{Im}\,\widehat{G}^R_\ell(\omega) = \frac{\pi \mathcal{A}}{E_b} \, \big[ F_\ell(q+\varpi) - F_\ell(q-\varpi) \big] = \frac{4\pi \mathcal{A}}{E_b^2} \, F'_\ell(q) \, \omega + O(\omega^3), \quad q>0
\end{equation}
where we have used \eqref{eq:Omega}.
Combining the two cases and using \eqref{eq:QLN_def}, we obtain the following expression for the quantum dissipative Love numbers:
\begin{equation}
\label{eq:nu_ell}
    \nu_\ell = \frac{\ell! \, (\ell+1)!} {2 \pi^3 \, \Gamma(2\ell+1) \, \Gamma(2\ell+3)} \, (r_+ E_b)^{2\ell+1} \, k \, F'_\ell(q) \, \Theta(q) \, , \qquad q \neq 0 \, ,
\end{equation}
where\footnote{Using the conjugation property $\psi(z^\ast)=\psi(z)^\ast$, the sum of digamma functions in \eqref{eq:fp_1} satisfies
\begin{equation*}
    i\sum_{\rho=\pm1}\sum_{\tau=\pm1}\tau\,\psi(\ell+1+i\rho k+i\tau\sqrt{q}) =-2\,\text{Im}\Big\{\psi[\ell+1+i(\sqrt{q}+k)]+\psi[\ell+1+i(\sqrt{q}-k)]\Big\} \, ,
\end{equation*}
which makes the reality of \eqref{eq:fp_1} manifest.}
\begin{equation}
\label{eq:fp_1}
    F'_\ell(q)
    =
    \frac{F_\ell(q)}{2 \sqrt{q}}
    \left[
        2 \pi \, \coth\!\left(2 \pi \sqrt{q}\right)
        +
        i \sum_{\rho=\pm1}\sum_{\tau=\pm1}
        \tau \,
        \psi\!\left(\ell+1+i\rho k+i\tau\sqrt{q}\right)
    \right] \, .
\end{equation}
To compare with the classical case of the dissipative Love numbers, we consider $\ell=0$ and $\ell=1$ as in \eqref{eq:nu_cl} where we find\footnote{For integer $\ell\ge1$, it is convenient to use the expression for $F'_\ell(q)$ given by
\begin{align*}
    F'_\ell(q)= \pi^2 \Bigg\{& Q'_\ell(q)\left[\coth \pi(\sqrt{q}-k)+\coth \pi(\sqrt{q}+k)\right]
    -\frac{\pi Q_\ell(q)}{2\sqrt{q}} \left[\text{csch}^2 \pi(\sqrt{q}-k)+\text{csch}^2 \pi(\sqrt{q}+k)\right]\Bigg\} \, .
\end{align*}}
\begin{align}
\label{eq:nu0}
    \nu_0 &= \frac{r_+ E_b}{4\pi}\left[k\coth(2\pi k)+\frac{1}{2\pi}\right] \, ,
    \\
    \nu_1 &= \Theta(q)\,\frac{(r_+E_b)^3}{48\pi}\,k\Bigg[(4k^2+5)\left(\coth\pi(\sqrt q-k)+\coth\pi(\sqrt q+k)\right)
    \nonumber\\
    &\hspace{27mm}
    +\frac{\pi(4k^2+1)}{\sqrt q}\left(\text{csch}^2\pi(\sqrt q-k)+\text{csch}^2\pi(\sqrt q+k)\right)\Bigg] \, .
    \label{eq:nu1}
\end{align}
More generally, using \eqref{eq:CDLN_def}, \eqref{eq:beta_NE} and \eqref{eq:nu_ell}, we obtain the ratio between the quantum dissipative Love numbers and their classical prediction:
\begin{equation}
\label{eq:nu_ratio}
    \frac{\nu_\ell}{\nu^\text{cl}_\ell}
    =
    \frac{F'_\ell(q) \, \Theta(q)}
    {2^{2 \ell} \pi^2 \Gamma(\ell + 1)^2 \, k^{2 \ell}}
    \, ,
    \qquad q \neq 0 \, .
\end{equation}
Here the comparison is between the low-frequency coefficients of the imaginary part of the quantum and classical responses. These quantum dissipative coefficients are determined by the scheme-independent imaginary response so their large $E/E_b$ limit can be compared directly with the corresponding classical coefficients. 
We show how the ratio in \eqref{eq:nu_ratio} behaves in Figure~\ref{fig:nu0_nu1} as a function of $E/E_b$ for the particular cases $\ell=0$ and $\ell=1$. As expected, in both cases the ratio approaches unity in the classical regime $E/E_b\gg1$. What may be less expected is the divergence for $E - E_{0,\ell} \to 0^+$. It comes from the thresholds of \eqref{eq:Im_GR_FT} at $\omega=\pm(E-E_{0,\ell})$, which approach $\omega=0$ in that limit. For $\ell = 1$, $\nu_1$ vanishes identically inside the gap $0<E<E_b$. We explain these phenomena in more detail below.

\subsection{Relation to other quantities}

\paragraph{Absorption and threshold behavior.} The imaginary part of the response function is related to the absorption cross-section $\sigma_\text{abs}^{(\ell)}(\omega)$ \cite{Emparan:2025sao,Biggs:2025nzs,Emparan:2025qqf} via 
\begin{equation}
    \sigma_\text{abs}^{(\ell)}(\omega) \propto \omega^{2 \ell - 1} \, \text{Im}\,\hat{\chi}_\ell(\omega) \, ,
\end{equation}
where the constant of proportionality is independent of $\omega$. Hence the dissipative Love numbers computed in Section~\ref{sec:QDLN} correspond to the leading coefficient in the small frequency expansion of the absorption cross-section
\begin{equation}
\label{eq:nu_ell_sigma_abs}
    \nu_\ell = \lim_{\omega \to 0} \frac{\text{Im}\,\hat{\chi}_\ell(\omega)}{\beta \omega} \propto \lim_{\omega \to 0} \frac{\sigma_\text{abs}^{(\ell)}(\omega)}{\beta \, \omega^{2 \ell}} \, .
\end{equation}
For $\ell \ge 1$, the dissipative Love numbers $\nu_\ell$ vanish for $0<E<E_{0,\ell}$ because of the step function $\Theta(q) = \Theta(E - E_{0,\ell})$ in \eqref{eq:nu_ell}. This comes from the spin-induced gap $E_{0,\ell}$ in the density of states \eqref{eq:rho_j}, and it is the low-frequency signature of the quantum transparency effect discussed in \cite{Emparan:2025qqf} for absorption cross-sections: $\sigma_\text{abs}^{(\ell)}(\omega)$ vanishes when $E + \omega < E_{0,\ell}$.

On the other hand, the divergence of the ratio $\nu_\ell / \nu^\text{cl}_\ell$ in \eqref{eq:nu_ratio} as $E - E_{0,\ell} \to 0^+$ corresponds to the enhancement of the low-frequency absorption cross-section \cite{Emparan:2025sao,Biggs:2025nzs,Emparan:2025qqf}, which explicitly takes the form
\begin{equation}
\label{eq:nu_ratio_small_q}
    \frac{\nu_\ell}{\nu^\text{cl}_\ell}
    \sim
    \frac{\left|\Gamma\!\left(\ell + 1 + i \sqrt{\ell (\ell + 1)} \right) \right|^4}
    {2^{2 \ell} \pi \Gamma(\ell + 1)^2 \left[\ell (\ell + 1) \right]^\ell}
    \, \sqrt{\frac{E_b}{2(E-E_{0,\ell})}} \, ,
\end{equation}
where the $\ell=0$ case is understood by continuity, and we have imposed the explicit value of $q$ via \eqref{eq:sA_p_q_Omega}. Thus, as the black hole energy approaches the threshold from above, the ratio $\nu_\ell / \nu^\text{cl}_\ell$ diverges like $1/\sqrt{E-E_{0,\ell}}$. This divergence can be understood from the square-root behavior of the density of states \eqref{eq:rho_j} near the threshold, $\rho_\ell(E)\propto\sqrt{E-E_{0,\ell}}$ for $E-E_{0,\ell} \to 0^+$. Indeed, this implies that the derivative $F'_\ell(q)$ appearing in \eqref{eq:nu_ell} behaves as $(E-E_{0,\ell})^{-1/2}$, leading to a divergence of $\nu_\ell$ for $\ell\geq1$ as the threshold is approached.\footnote{For $\ell=0$, for which $E_{0,\ell} = 0$, the factor $1/\beta\propto\sqrt{E}$ in the definition \eqref{eq:QLN_def} cancels the $E^{-1/2}$ divergence of $F'_0(q)$, so that $\nu_0$ approaches a finite value as $E \to 0$. For $\ell\ge1$, instead, $\nu_\ell$ itself diverges as $(E - E_{0,\ell})^{-1/2}$. On the other hand, for any $\ell \ge 0$, the ratio $\nu_\ell / \nu^\text{cl}_\ell$ diverges as $(E - E_{0,\ell})^{-1/2}$ as shown in \eqref{eq:nu_ratio_small_q}.}

The same square-root behavior of \eqref{eq:rho_j} produces branch points in the quantum response function $\hat{\chi}_\ell(\omega)$ \cite{Emparan:2026ta}. These occur at
\begin{equation}
\label{eq:branch_points}
    \text{branch points: } \qquad \omega = \pm (E - E_{0,\ell}) \, .
\end{equation}
For $E>E_{0,\ell}$, these branch points lie away from $\omega=0$. Close to the threshold, they set the scale controlling the low-frequency expansion \eqref{eq:chi_qu_series_LN}, which requires
\begin{equation}
    \frac{|\omega|}{E-E_{0,\ell}} \ll 1 \, .
\end{equation}
However, as $E-E_{0,\ell}\to0^+$, the branch points approach $\omega=0$, and the range of validity of the low-frequency expansion shrinks to zero. At the exact threshold $E=E_{0,\ell}$, the two branch points coincide at $\omega=0$ and the low-frequency behavior of the imaginary part \eqref{eq:Im_GR_FT} of the quantum response function is
\begin{equation}
    \text{Im}\,\hat{\chi}_\ell(\omega)
    \propto
    \text{sign}(\omega)\sqrt{|\omega|} \, .
\end{equation}
Thus the response is continuous but non-analytic at $\omega=0$. Hence it cannot be characterized by a dissipative Love number defined through a low-frequency Taylor expansion in $\omega$. For $\ell \ge 1$, the divergence of $\nu_\ell$ as $E - E_{0,\ell} \to 0^+$ and the loss of analyticity at the exact threshold therefore have the same origin coming from the square-root behavior of the Schwarzian density of states. The corresponding branch point is a genuine nonlocal feature of the retarded Green's function and cannot be modified by local counterterms.\footnote{The finiteness of $\nu_0$ as $E \to 0$ does not restore analyticity at the threshold $E_{0,\ell=0} = 0$: at $E = 0$ the response still behaves as $\text{Im}\,\hat{\chi}_0(\omega)\propto\text{sign}(\omega)\sqrt{|\omega|}$ and hence does not admit a Taylor expansion around $\omega=0$.}

In the time domain, this is probing the late-time Schwarzian tail $\sim t^{-3/2}$, rather than a short-memory quasinormal decay. This is because the branch-cut contribution to the time-domain response function is of the form $\chi_\ell(t) \propto (E_b t)^{-3/2} \sin\!\left[(E - E_{0,\ell}) t - 3 \pi/4 \right]$ \cite{Emparan:2026ta}. As the threshold is approached, the oscillation timescale $(E-E_{0,\ell})^{-1}$ becomes large, and $\nu_\ell / \nu^\text{cl}_\ell$ becomes increasingly sensitive to the late-time Schwarzian tail. The divergence of $\nu_\ell / \nu^\text{cl}_\ell$ is therefore a consequence of the long-lived Schwarzian throat fluctuations, not a sign of a linear instability: it arises from a real-frequency branch point rather than from a pole in the upper half-plane. Its consequences for absorption rates must be inferred from the full non-analytic response. At the exact $E = E_{0,\ell}$ threshold, the absorption cross-section scales as $\sigma_\text{abs}^{(\ell)}(\omega) \propto\omega^{2\ell-1/2}$. Therefore, as $\omega \to 0$ the absorption cross-section at the gap threshold is singular for $\ell = 0$. On the other hand, it vanishes for $\ell \ge 1$, although the corresponding dissipative coefficient $\nu_\ell$ diverges as $E - E_{0,\ell} \to 0^+$.

These results require that the states involved are far enough from the bottom of their spectrum for the continuum density \eqref{eq:rho_j} to apply. The imaginary part \eqref{eq:Im_GR_FT} selects intermediate spin-$\ell$ states of energy $E'=E\pm\omega$, which approach their spectral edge $E_{0,\ell}$ when both $E-E_{0,\ell}$ and $\omega$ are small. For $\ell\geq1$ this is distinct from the limit $E\to0^+$, where it is the initial spinless black hole that approaches its own edge. In pure JT gravity, the disk approximation that we have used breaks down at energies of order $e^{-2S_0/3}E_b$ above the edge \cite{Saad:2019lba,Iliesiu:2019lfc,Mertens:2022irh}. For a four-dimensional black hole, non-perturbative contributions not captured by JT gravity may become relevant at a different scale \cite{Iliesiu:2020qvm}, which has not been determined. The threshold behavior found above, namely the growth of $\nu_\ell$, the $\sqrt{|\omega|}$ response and the $t^{-3/2}$ tail, holds between this scale and $E_b$, but not in the strict limits $E\to E_{0,\ell}^+$ and $E\to0^+$.

\paragraph{Transport coefficients.}
Several recent works have studied Schwarzian corrections to shear viscosity and electrical conductivity in near-extremal black branes
in the context of holographic transport, extracting these coefficients from the low-frequency dissipative response through Kubo formulas
\cite{Nian:2025oei,PandoZayas:2025snm,Cremonini:2025yqe,Gouteraux:2025exs,Kanargias:2025vul,Maulik:2026dro,Das:2026ujn}.
This provides a natural comparison with the dissipative Love numbers defined in \eqref{eq:QLN_def}: in both cases, the observable is determined, up to normalization, by the coefficient linear in frequency in the imaginary part of a retarded Green's function, whenever this expansion
exists. In the holographic setting, the response of an asymptotically $\text{AdS}$ black brane determines a transport coefficient of the dual field theory. In our case, the response characterizes dissipation by an asymptotically flat black hole, and the corresponding coefficient is the dissipative Love number.

In general, what we find is that the same long-lived Schwarzian throat dynamics responsible for the growth of $\nu_\ell / \nu^\text{cl}_\ell$ near extremality also enhances transport coefficients in related near-extremal black brane settings, such as shear viscosity and electrical conductivity, which we now describe.

The shear viscosity $\eta$ of the fluid dual to a near-extremal black brane is obtained from the low-frequency absorption of a shear graviton $h_{xy}$ with zero momentum along the brane. This perturbation obeys the equation of a minimally coupled massless scalar, and in the $\text{AdS}_2$ throat it reduces to a homogeneous scalar mode with $\Delta = 1$, the same as our s-wave scalar perturbation. Since $\eta$ measures the dissipative response of this mode, its natural counterpart in our setup is the dissipative Love number $\nu_0$.
The quantum-to-classical ratios $\eta / \eta^\text{cl}$ and $\nu_0 / \nu^\text{cl}_0$ show a qualitatively similar behavior: they are enhanced in the quantum Schwarzian regime, diverge as the low-energy threshold $E \to 0$ is approached, and tend to unity in the classical regime $E/E_b \gg 1$ \cite{Nian:2025oei,PandoZayas:2025snm,Cremonini:2025yqe,Gouteraux:2025exs,Kanargias:2025vul}.\footnote{This comparison is only qualitative. Our analysis is microcanonical, whereas the viscosity is computed in the canonical ensemble, and in the quantum Schwarzian regime the two ensembles are not equivalent, unlike in the semiclassical regime \cite{Emparan:2025sao,Biggs:2025nzs,Biggs:2026zlp}. Our expression \eqref{eq:nu0} for $\nu_0$ therefore cannot be mapped to the viscosity results of \cite{Nian:2025oei,PandoZayas:2025snm,Cremonini:2025yqe,Gouteraux:2025exs,Kanargias:2025vul} by the relation \eqref{eq:beta_NE} between temperature and energy.} 

Likewise, the electrical conductivity is associated with the homogeneous $\text{AdS}_2$ scalar mode with $\Delta = 2$ descending from a four-dimensional transverse photon perturbation with zero momentum. In this case, the analogy with the $\ell = 1$ scalar perturbation in our setup is only partial: although both modes are associated with $\Delta = 2$, our $\ell = 1$ scalar carries nonzero angular momentum on the transverse $S^2$ and changes the spin of the black hole. As a consequence, the corresponding dissipative Love number $\nu_1$ is sensitive to the spin-one threshold in \eqref{eq:gap}. By contrast, the photon mode relevant for the conductivity is homogeneous along the transverse $\mathbb{R}^2$ or $T^2$ directions of the black brane and does not involve an analogous threshold.
Nonetheless, the electrical conductivity was also found to increase at sufficiently low temperatures in the quantum Schwarzian regime \cite{Maulik:2026dro}. 

In \cite{Kanargias:2025vul}, the divergence of the shear viscosity was interpreted as a possible indication of glassy low-energy dynamics. In the present setting, the interpretation should be more conservative, since $\nu_\ell$ is not a transport coefficient of a dual boundary fluid, and our calculation is microcanonical and asymptotically flat. Nevertheless, the analogy is suggestive: in both cases, a Schwarzian soft mode controls the low-frequency dissipative response and enhances it near extremality. From this point of view, the enhancement of the dissipative Love numbers relative to their classical values reflects the increasing sensitivity of the quantum near-$\text{AdS}_2$ throat to very soft perturbations.

The same enhancement can also be understood in the time domain. As discussed above, the branch-point structure responsible for the singularity in the low-frequency coefficients is associated with the late-time power-law tail of the Schwarzian response. Thus, as the threshold is approached, the low-frequency dissipative response becomes increasingly sensitive to long timescales in the near-$\text{AdS}_2$ throat. This provides another way of viewing the analogy with glassy systems: the comparison is not with a literal hydrodynamic glass phase, but with an infrared response dominated by increasingly slow collective dynamics.

\paragraph{Black hole-induced decoherence.} At the semiclassical level, black holes are known to decohere quantum superpositions in their exterior \cite{Danielson:2022tdw,Danielson:2022sga,Gralla:2023oya}. In the semiclassical thermodynamic regime, the microcanonical and canonical ensembles are equivalent, so one may use the canonical correlators, which satisfy the KMS condition. The fluctuation-dissipation theorem then relates the zero-frequency Wightman correlator entering the decoherence rate to the low-frequency imaginary part of the response function \cite{Biggs:2024dgp}. Consequently, at leading order and up to normalization factors determined by the external source, the decoherence rate induced by the black hole is controlled by the corresponding dissipative Love number.

The quantum microcanonical regime considered here is different \cite{Li:2025vcm,Biggs:2026zlp}. As emphasized in \cite{Biggs:2026zlp}, while the decoherence rate is governed by the zero-frequency Wightman function, the dissipative Love number is extracted from the low-frequency imaginary part of the retarded Green's function. In the quantum Schwarzian regime, the microcanonical and canonical ensembles are no longer equivalent, and the fluctuation-dissipation relation between the Wightman and retarded correlators does not apply. Therefore, the proportionality between the decoherence rate and the dissipative Love number is a semiclassical relation; in the quantum microcanonical regime considered here, the two quantities are distinct observables and cannot in general be inferred from one another.

\section{Conclusions}
\label{sec:conclusions}

Close to extremality the throat of a black hole no longer has a definite classical geometry. It responds to external fields as any quantum system does: through transitions among its states. We have used this to define and compute its response to a scalar field in the regime where the Schwarzian mode is strongly coupled. The static response is a polarizability, and for $\ell\geq1$ it does not vanish, since the rotational energy of the spin-$\ell$ intermediate states spoils the cancellation that holds classically. For $\ell=0$, instead, the shift symmetry of the scalar protects the monopole and forces its static Love number to vanish exactly. 

The near-horizon theory fixes how the physical conservative response depends on the energy and the frequency up to local terms.
For $\ell\geq1$ this dependence is logarithmic when the frequency or $\sqrt{2E_bE}$ exceeds the rotational energy $E_{0,\ell}$, with the coefficient fixed by the running; otherwise there is no logarithm of the energy nor the frequency, but the dependence remains non-polynomial.
The dissipative response is instead fixed completely: below the rotational threshold it vanishes at sufficiently low frequency, and at the threshold it is non-analytic. Table~\ref{tab:love_numbers} collects the expressions for the static and dissipative Love numbers.

\begin{table}[h!]
    \centering
    \begin{tabular}{|c|c|c|c|c|}
        \hline
        \multicolumn{2}{|c|}{Love numbers}
        & classical & quantum & comments \\ \hline
        static & $\lambda_{\ell=0}$
        & 0~\eqref{eq:lambda_cl_zero} &  ~~0~\eqref{eq:shift_Ward_GR}
        & vanishes by shift symmetry \\ \hline
        static & $\lambda_{\ell\geq 1}$
        & 0~\eqref{eq:lambda_cl_zero}
        & $\neq 0$~\eqref{eq:lambda_ell_scales}
        & \begin{tabular}[c]{@{}c@{}}
            requires renormalization;\\
            nonzero beta function
          \end{tabular} \\ \hline
        dissipative & $\nu_{\ell\geq 0}$
        & \eqref{eq:CDLN_def}
        & \eqref{eq:nu_ell}
        & does not require renormalization \\ \hline
    \end{tabular}
    \caption{\small Static and dissipative Love numbers for a massless
    neutral scalar field. Equation references give the corresponding
    expressions.}
    \label{tab:love_numbers}
\end{table}

The formulation through a renormalized generating functional is what makes this analysis possible. Classically it reproduces the Love numbers extracted from radial coefficients, as we showed in Section~\ref{sec:CLN}, and it remains meaningful when there is no radial profile to read them from. It also shows where the quantum ambiguity comes from. In the classical throat the local terms are built from $\partial_t^2$ and $E_bE$, both of dimension two, while the response has odd dimension, so none of them is finite. The loops of the Schwarzian mode bring the scale $E_b$, of dimension one, and with it the logarithm and the finite local terms.

These finite local terms are the main limitation of our analysis: the near-horizon theory does not in general fix them. They encode the physics at the mouth of the throat and beyond it; fixing them requires the four-dimensional effective theory, including its loop corrections and higher-derivative couplings \cite{Noumi:2026shc}. We have attempted neither a systematic classification of these terms nor their matching to the four-dimensional theory. This bears on the size of the effect: with the conventional normalization by the horizon radius, the throat contribution is parametrically small, and the loop corrections that do not rely on the throat, whose running was found in \cite{Elkhidir:2026lgp}, are expected to be larger. What the throat contributes, and what those corrections cannot mimic, is a non-analytic dependence on the energy above extremality at the scale $E_b$.

Other restrictions are less fundamental: we have taken the black hole to be initially spinless and at fixed energy, we have kept only the leading order in $r_+\omega$, and our expressions cease to apply very close to the threshold, where the spectrum is no longer a smooth continuum.

We have studied a scalar probe, but the mechanism is not specific to it. Any perturbation with $\ell\geq1$ transfers angular momentum to the black hole, and the rotational energy enters the energy differences in the same way. We therefore expect the electromagnetic and gravitational Love numbers, which in the throat are also governed by operators of integer dimension, to become nonzero for the same reason. Their calculation requires the coupled perturbations of the metric and the gauge field, and it has no counterpart of the shift symmetry that protects the scalar monopole.

Two other settings seem to us especially interesting. The first is the near-extremal Kerr black hole, where the rotational mode is of a different kind and the perturbations can extract angular momentum as well as supply it. There, however, superradiant emission is not a small effect in the quantum regime, and whether a quasi-stationary response can be defined at all requires care. The second is that of supersymmetric black holes, whose ground states are separated from the continuum by a gap already in the Schwarzian sector \cite{Heydeman:2020hhw}: the static response of such states would be a polarizability of a different origin from the one found here.

\section*{Acknowledgments}

The work of PAC is supported by a Ramón y Cajal fellowship (RYC2023-044375-I) and by a Proyecto de Generaci\'on de Conocimiento (PID2024-155685NB-C22) from Spain’s Ministry of Science, Innovation and Universities. RE and ST are supported by MICIU grants PID2022-136224NB-C22 and PID2025-169343NB-C21, and grant CEX2024-001451-M funded by MICIU/AEI/10.13039/501100011033. The project that gave rise to these results received the support of a fellowship from ``la Caixa” Foundation (ID 100010434), awarded to ST, with code LCF/BQ/DI24/12070002. MD is funded by the Postdoctoral Fellows of the Research Foundation - Flanders grant (1235324N) and by the European Union’s Horizon Europe research and innovation programme under the Marie Skłodowska-Curie grant agreement No. 101206221. This research was carried out in part during a scientific visit of MD to ICCUB funded by the COST action CA22113 ``Fundamental Challenges in Theoretical Physics." MD extends profound gratitude to the ICCUB group for establishing an exceptional benchmark of hospitality, thereby ensuring all subsequent academic visits will pale in comparison.

\appendix
\section{Classical static Love numbers in $D$ dimensions}
\label{app:CLN_Ddim}

In Section~\ref{sec:CLN}, we computed the classical scalar response of a four-dimensional near-extremal Reissner--Nordstr\"{o}m black hole in the low-frequency regime. In the strictly static limit $\omega=0$, however, the scalar wave equation can be solved exactly throughout the full geometry, without focusing on the near-horizon. In this appendix, we perform this calculation in $D=n+3$ spacetime dimensions and compute the corresponding classical static Love numbers, recovering the known expressions \cite{Pereniguez:2021xcj,Charalambous:2024tdj}. For $D=4$, the result provides an independent check of the static response obtained in Section~\ref{sec:CLN}.

The $D=n+3$ dimensional Reissner--Nordstr\"{o}m metric is
\begin{equation}
    ds^2 = - f(r) \, dt^2 + \frac{dr^2}{f(r)} + r^2 \, ds^2_{S^{n+1}} \, ,
\end{equation}
where
\begin{equation}
    f(r) = 1 - \frac{2 m}{r^n} + \frac{q^2}{r^{2 n}} = \frac{(r^n - r_+^n) \, (r^n - r_-^n)}{r^{2 n}} \, , \qquad r_\pm^n = m \pm \sqrt{m^2 - q^2} \, ,
\end{equation}
and $ds^2_{S^{n+1}}$ is the metric on the unit $(n+1)$-sphere. The parameters $m$ and $q$ are related to the ADM mass $M$ and charge $Q$ of the black hole, while $r=r_{\pm}$ are the outer and inner horizons. The inverse Hawking temperature is
\begin{equation}
    \beta = \frac{4 \pi r_+^{n+1}}{n \, (r_+^n - r_-^n)} = \frac{4 \pi r_+}{n \varepsilon} \, ,
\end{equation}
where
\begin{equation}
\label{eq:xi_def_D}
    \varepsilon = \frac{r_+^n - r_-^n}{r_+^n} \, .
\end{equation}
Thus $\varepsilon$ measures the departure from extremality, with $\varepsilon\to0$ as $r_-\to r_+$. For $D=4$, where $n=1$, \eqref{eq:xi_def_D} reduces to the parameter $\varepsilon$ defined in \eqref{eq:xi}. As in Section~\ref{sec:CLN}, we consider a minimally coupled, neutral, massless scalar field $\Phi = \Phi(t,r,\theta_1,\ldots,\theta_{n+1})$ with action
\begin{equation}
    I = -\frac{1}{2} \, \int d^D x \, \sqrt{-g} \, \nabla_\mu \Phi \, \nabla^\mu \Phi \, .
\end{equation}
Its equation of motion in the black hole background is
\begin{equation}
\label{eq:Phi_EOM_D}
    \Box_\text{\tiny BH} \Phi = 0 \, , \qquad \Box_\text{\tiny BH} = \nabla_\mu \nabla^\mu = - \frac{1}{f} \, \partial_t^2 + \frac{1}{r^{n+1}} \, \partial_r(r^{n+1} \, f \, \partial_r) + \frac{1}{r^2} \, \nabla^2_{S^{n+1}} \, ,
\end{equation}
where $\nabla^2_{S^{n+1}}$ is the Laplace--Beltrami operator on $S^{n+1}$. Since we are interested in the static response, we take $\Phi$ to be time-independent and decompose it as
\begin{equation}
    \Phi(r,\theta_1,\ldots,\theta_{n+1}) = \psi(r)\, Y_\ell(\theta_1,\ldots,\theta_{n+1}) \, ,
\end{equation}
where we suppress the degeneracy labels of the spherical harmonics and use
\begin{equation}
    \nabla^2_{S^{n+1}}Y_\ell = -\ell(\ell+n)Y_\ell \, .
\end{equation}
Inserting this ansatz into the equation of motion \eqref{eq:Phi_EOM_D}, we obtain the radial equation
\begin{equation}
\label{eq:5_radEOM}
    \frac{1}{r^{n-1}} \, \partial_r(r^{n+1} \, f \, \psi') - \ell \, (\ell + n) \, \psi = 0 \, .
\end{equation}
Equation~\eqref{eq:5_radEOM} can be brought to Legendre form and solved exactly:
\begin{equation}
\label{eq:5_FullSol}
    \psi(r) = C_1 \, P_{\ell / n}\left(1 + \frac{2 (r^n - r_+^n)}{\varepsilon r_+^n} \right) + C_2 \, Q_{\ell / n}\left(1 + \frac{2 (r^n - r_+^n)}{\varepsilon r_+^n} \right) \, .
\end{equation}
Regularity at the horizon requires that
\begin{equation}
\label{eq:hor_reg}
    C_2 = 0 \, .
\end{equation}
To extract the static response, we expand the solution at large $r$. In direct analogy with \eqref{eq:IZ_sol}, we write
\begin{equation}
    \psi(r)
    =
    A\left(\frac{r}{r_+}\right)^\ell
    \big(1+\cdots\big)
    +
    B\left(\frac{r}{r_+}\right)^{-(\ell+n)}
    \big(1+\cdots\big) \, ,
\end{equation}
where the first term is the applied source and the second is the induced response. For generic values of $\ell/n$, the large-$r$ expansion of \eqref{eq:5_FullSol} gives
\begin{equation}
    A = C_1 \, \frac{\Gamma\!\left(\frac{2 \ell}{n} + 1 \right)}{\Gamma\!\left(\frac{\ell}{n} + 1 \right)^2} \, \varepsilon^{-\frac{\ell}{n}} \, ,\qquad B = C_1 \, \frac{\Gamma\!\left(-\frac{2 \ell}{n} - 1 \right)}{\Gamma\!\left(-\frac{\ell}{n} \right)^2} \, \varepsilon^{\frac{\ell}{n} + 1} \, .
\end{equation}
Using the same normalization as in \eqref{eq:chi_def}, the classical static Love numbers are therefore
\begin{align}
\label{eq:lambda_cl_RN}
    \lambda_\ell^\text{cl} = \frac{B}{2 A}
    = \frac{\varepsilon^{\frac{2 \ell}{n} + 1}}{2} \, \frac{\Gamma\!\left(\frac{\ell}{n} + 1 \right)^2 \, \Gamma\!\left(-\frac{2 \ell}{n} - 1 \right)}{\Gamma\!\left(\frac{2 \ell}{n} + 1 \right) \, \Gamma\!\left(-\frac{\ell}{n} \right)^2} \, ,
\end{align}
in agreement with the known expressions in the literature \cite{Pereniguez:2021xcj,Charalambous:2024tdj}. For generic $\ell/n$, \eqref{eq:lambda_cl_RN} can be evaluated directly. Special care is required when $2\ell/n\in\mathbb{Z}$, for which the two large-radius branches of the Legendre equation become resonant. There are two qualitatively different cases. If $\ell/n\in\mathbb{Z}$, equation \eqref{eq:lambda_cl_RN} is defined by analytically continuing in $\ell/n$ and then taking the integer limit, which gives
\begin{equation}
    \lambda_\ell^\text{cl}=0 \, , \qquad \frac{\ell}{n}\in\mathbb{Z} \, .
\end{equation}
In particular, for $D=4$ we have $n=1$. Using $\varepsilon=4\pi r_+/\beta$, the analytically continued expression \eqref{eq:lambda_cl_RN} reproduces the static Love number \eqref{eq:CSLN_def} obtained in Section~\ref{sec:CLN}, and therefore vanishes for every integer $\ell$, in agreement with \eqref{eq:lambda_cl_zero}.\footnote{The vanishing of the classical Love numbers of four-dimensional black holes admits several symmetry-based interpretations, including ladder symmetries of the static perturbation equations and the near-zone Love symmetry \cite{Charalambous:2021kcz,Hui:2021vcv,BenAchour:2022uqo,Charalambous:2022rre,Berens:2022ebl,Rai:2024lho}.}

If instead $\ell/n$ is half-integer, the horizon-regular solution contains a logarithmic term in its large-radius expansion and the static response requires renormalization. After subtracting the pole in \eqref{eq:lambda_cl_RN}, the finite Love number depends on the renormalization scale, while the residue of the pole determines the coefficient of the logarithm and hence the corresponding beta function. This classical running of the static Love numbers in $D \ge 5$ physically means that their value depends on the radial distance at which the response of the system is measured \cite{Hui:2020xxx}. This is different from the quantum running of the $D=4$ static Love numbers in the quantum Schwarzian regime discussed in Section~\ref{sec:QSLN}.

\section{Counterterms for the classical response}
\label{app:ct}

We work out the holographic renormalization of Section~\ref{subsec:retarted2pointfunction} for $\ell=1$, the simplest case in which a divergence depends on the energy of the black hole.

In terms of the coordinate $\zeta$ of \eqref{eq:zeta_def}, and with $(2\pi/\beta)^2 = 2 E_b E$ from \eqref{eq:beta_NE}, the blackening factor of the throat \eqref{eq:NH_metric_Z} is $1 - Z^2 = 1 - 2 E_b E \, \zeta^2$. We cut off the throat at $\zeta = \zeta_0$, where the induced metric is $\sqrt{-\gamma} = (r_+/\zeta_0) \sqrt{1 - 2 E_b E \, \zeta_0^2}$. On-shell, the action \eqref{eq:I_bulk_cl_aux} reduces to boundary terms, and under a change of the solution, it varies at the cutoff by
\begin{equation}
\label{eq:dI_bulk}
    \delta I_\text{bulk} \Big|_{\zeta_0} = r_+^2 \sum_{\ell,m} \int dt \; \delta\Psi_{\ell m}^\ast \left( 1 - 2 E_b E \, \zeta^2 \right) \partial_\zeta \Psi_{\ell m} \Big|_{\zeta_0} \, .
\end{equation}
For $\ell=1$ and frequency $\omega$, the solution of \eqref{eq:radialmaster_scaling} near the boundary is
\begin{equation}
    \psi = A \, \frac{r_+}{\zeta} \left[ 1 + \frac{\omega^2 \zeta^2}{2} + O(\zeta^4) \right] + B \, \frac{\zeta^2}{r_+^2} \left[ 1 + O(\zeta^2) \right] ,
\end{equation}
and inserting it in \eqref{eq:dI_bulk} we find
\begin{equation}
    \delta I_\text{bulk} \Big|_{\zeta_0} = r_+^4 \left( - \frac{1}{\zeta_0^3} + \frac{2 E_b E}{\zeta_0} \right) A \, \delta A^\ast + r_+ \left( 2 B \, \delta A^\ast - A \, \delta B^\ast \right) + O(\zeta_0) \, .
\end{equation}
The second divergence comes from the blackening factor in \eqref{eq:dI_bulk}. Both are removed by
\begin{equation}
\label{eq:Ict_bulk}
    I_\text{ct} = \frac{r_+}{2} \sum_{m} \int dt \, \sqrt{-\gamma} \; \Psi_{1 m}^\ast \left[ 1 - \hat\zeta_0^{\,2} \left( -\partial_t^2 + E_b H \right) \right] \Psi_{1 m} \, , \qquad \hat\zeta_0 \equiv \frac{r_+}{\sqrt{-\gamma}} \, ,
\end{equation}
where $\hat\zeta_0$ is the cutoff in units of proper time and $H$ takes the value $E$. Its variation is
\begin{equation}
    \delta I_\text{ct} = r_+^4 \left( \frac{1}{\zeta_0^3} - \frac{2 E_b E}{\zeta_0} \right) A \, \delta A^\ast + r_+ \left( B \, \delta A^\ast + A \, \delta B^\ast \right) + O(\zeta_0) \, ,
\end{equation}
so that
\begin{equation}
    \delta \left( I_\text{bulk} + I_\text{ct} \right) \Big|_{\zeta_0} \; \xrightarrow[\zeta_0\to 0]{} \; 3 \, r_+ \, B \, \delta A^\ast \, ,
\end{equation}
which is \eqref{eq:deltaIbulkct}. The terms with $\delta B^\ast$ cancel, as they must for the variation to be with respect to the source only.

The first term in \eqref{eq:Ict_bulk} is the standard one. In the others the Hamiltonian enters on the same footing as the time derivatives; without it, a divergence $r_+^4 E_b E \, A \, \delta A^\ast/\zeta_0$ would remain. These are the terms of \eqref{eq:source_alpha} with $(\alpha,c) = (0,0)$, $(2,0)$ and $(0,1)$. On the cutoff surface the energy is not carried by the induced metric, which for a line has no curvature, but by the extrinsic curvature $K$ of the surface, 
\begin{equation}
    (r_+ K)^2 = 1 + 2 \hat\zeta_0^{\,2} E_b E\,. 
\end{equation}
For higher $\ell$, the bracket in \eqref{eq:Ict_bulk} is a polynomial of degree $\ell$ in $\hat\zeta_0^{\,2} \partial_t^2$ and $\hat\zeta_0^{\,2} E_b H$. For $\ell = 0$ there is no divergence at all.

\section{Classical limit at integer multipoles}
\label{app:class_lim} 

In Section~\ref{subsec:class_lim}, we recovered the full classical response \eqref{eq:chi_result} starting from the regularized expression \eqref{eq:GR_FT_SP}. To do so, we analytically continued $\ell$ to the convergence range \eqref{eq:ell_conv}, took the classical limit, and then continued back to integer values of $\ell$. While this procedure recovers the full analytic-in-$\ell$ classical result \eqref{eq:chi_result}, it hides the finite renormalization ambiguity parametrized by the polynomial $P_\ell(\omega^2,k^2)$ in \eqref{eq:Re_GR_ren}. Here we work directly at integer $\ell$ and show how the renormalized expression \eqref{eq:Re_GR_ren} reproduces the scheme-independent classical result \eqref{eq:Re_chi_int}. This imposes a matching condition on the classical asymptotics of $P_\ell(\omega^2,k^2)$.

The renormalized real part of the quantum response function \eqref{eq:Re_GR_ren}, which we rewrite here for clarity, is given by
\begin{align} \label{eq:Re_GR_ren_app_3}
    \text{Re}\,\widehat{G}^{R,\text{ren}}_\ell(\omega) &= \frac{ 2 \pi^2\mathcal{A}}{E_b} \sum_{\sigma = \pm 1} \left[ Q_\ell(q + \sigma \varpi) \log \frac{\mu}{|q + \sigma \varpi|} \nonumber
    \right. \\& \left. \qquad \qquad \qquad \qquad + \frac{1}{2\pi^2}\text{PV} \int_0^{+\infty} dp \frac{R_\ell(p)}{p-q-\sigma \varpi} \right]+ P_\ell(\omega^2,k^2) \, ,
\end{align}
where $\mathcal{A}$ is given in \eqref{eq:A-const}, the function $Q_{\ell}$ in \eqref{eq:Qell} and $q=k^2-\ell(\ell+1)$ and $\varpi=2k\tilde{\omega}$, are found in \eqref{eq:sA_p_q_Omega} and \eqref{eq:Omega} respectively.

The first term proportional to the logarithm vanishes in the classical limit \eqref{eq:class_lim}. To see this, we first note that $\mathcal{A}/E_b = O(k^{-2\ell-1})$ while $Q_\ell(q+\sigma\varpi)=O(k^{2\ell+1})$ and is odd under $\sigma\to-\sigma$. On the other hand, the $\sigma$-even part of the logarithm is $O(\log k)$, while its $\sigma$-odd part is $O(k^{-1})$. Putting it all together, we find
\begin{equation}
    \frac{2 \pi^2 \mathcal{A}}{E_b} \sum_{\sigma = \pm 1} Q_\ell(q + \sigma \varpi) \log \frac{\mu}{|q + \sigma \varpi|} = O\!\left(\frac{\log k}{k} \right) \, ,
\end{equation}
which vanishes for large $k$. The second term in \eqref{eq:Re_GR_ren_app_3} requires us to study the behavior of $R_\ell(p)$. To this end, we use the variable $s$ defined in \eqref{eq:ptosdefinition}, i.e., $s=\sqrt{p}-k$,
and we separate the discussion into the three cases $s \ll -1$, $s=O(1)$, and $s \gg 1$. For $s\ll-1$ or $s\gg1$, the ratio of hyperbolic functions in \eqref{eq:Rell} approaches $0$ or $2$ exponentially fast, respectively. Hence,
\begin{equation}
\label{eq:Rell_cl_large_s}
    R_\ell(p) \simeq -2 \pi^2 Q_\ell(p) \, \Theta(k^2 - p) \, , \qquad |s| \gg 1 \, .
\end{equation}
It is then convenient to decompose $R_\ell(p)$ as follows, for any value of $s$:
\begin{equation}
\label{eq:Rell_decomp}
    R_\ell(p) = \delta R_\ell(p) -2 \pi^2 Q_\ell(p) \, \Theta(k^2 - p) \, ,
\end{equation}
where
\begin{equation}
\label{eq:delta_R_explicit}
    \delta R_\ell(p) \equiv \pi^2 Q_\ell(p) \left[ \frac{\sinh(2\pi\sqrt{p})} {\sinh^2(\pi\sqrt{p})-\sinh^2(\pi k)} -2\Theta(p-k^2) \right] \, .
\end{equation}
Because of \eqref{eq:Rell_cl_large_s}, the remainder $\delta R_\ell(p)$ is exponentially suppressed for $|s|\gg1$ and is therefore localized around $s=O(1)$ in the classical limit.

We now insert the decomposition \eqref{eq:Rell_decomp} into the principal value integral in \eqref{eq:Re_GR_ren_app_3}. As we show below, the contribution coming from $\delta R_\ell(p)$ vanishes in the classical limit, while the second reduces to a local polynomial of degree $\ell$ in $\omega^2$, which can then be cancelled by the classical asymptotics of $P_\ell(\omega^2,k^2)$.

To analyze the contribution of $\delta R_\ell(p)$ to \eqref{eq:Re_GR_ren_app_3}, we first study its behavior in the classical limit at fixed $s = \mathcal{O}(1)$. Since
\begin{equation}
    Q_\ell\left((k+s)^2\right) = (2k)^{2\ell+1} s\prod_{m=1}^{\ell}(s^2+m^2) \left[1+O(k^{-1})\right]\,,
\end{equation}
and
\begin{equation}
    \frac{\sinh\!\left(2\pi(k+s)\right)}
    {\sinh^2\!\left(\pi(k+s)\right)-\sinh^2(\pi k)}
    = 1+\coth(\pi s)+O(e^{-4\pi k}) \, ,
\end{equation}
we find to leading order in large $k$
\begin{equation}
\label{eq:delta_f_class_asymp}
    \delta R_\ell\left((k+s)^2\right)
    = \pi^2(2k)^{2\ell+1} s
    \left(\coth(\pi s)-\text{sign}(s)\right)
    \prod_{m=1}^{\ell}(s^2+m^2)
    \left[1+O(k^{-1})\right] \, ,
\end{equation}
which, as expected, is exponentially suppressed as $|s|\to\infty$. The integration measure and the rational function in \eqref{eq:Re_GR_ren_app_3} are given, in the classical regime, by
\begin{equation}
\label{eq:class_kernel_scaling}
    dp = 2(k+s)\,ds \, , \qquad \frac{p-q}{(p-q)^2-\varpi^2} = \frac{1}{2k}\,\frac{s}{s^2-\tilde{\omega}^2} +O(k^{-2}) \, .
\end{equation}
Since $\delta R_\ell(p)$ is localized in the region $s=O(1)$, the integration range can be extended from $s\in[-k,+\infty)$ to the whole real line in the classical limit, and we obtain
\begin{align}
\label{eq:delta_f_GR_class}
    &\frac{2\mathcal{A}}{E_b}
    \,\text{PV}\int_0^{+\infty}dp\,
    \delta R_\ell(p)
    \frac{p-q}{(p-q)^2-\varpi^2}
    \nonumber\\
    &\qquad\stackrel{\text{cl}}{\longrightarrow}
    \frac{2\ell+1}{\Gamma(2\ell+2)}
    (E_bk)^{2\ell+1}
    \,\text{PV}\int_{-\infty}^{+\infty}ds\,
    \left[\coth(\pi s)-\text{sign}(s)\right]
    \frac{s^2}{s^2-\tilde{\omega}^2}
    \prod_{m=1}^{\ell}(s^2+m^2) \, .
\end{align}
The factor $(\coth(\pi s)-\text{sign}(s))$ is odd under $s\to-s$, whereas all the remaining factors in the integrand are even. The principal value prescription treats the pair of poles at $s=\pm\tilde{\omega}$ symmetrically and is therefore invariant under $s\to-s$. Therefore the integral in \eqref{eq:delta_f_GR_class} vanishes.

It remains to analyze the contribution of the second term in \eqref{eq:Rell_decomp}. Adding and subtracting $Q_\ell(q+\sigma\varpi)$ in the numerator, we obtain
\begin{align}
\label{eq:Q_Theta_cl}
    &\frac{2\pi^2\mathcal{A}}{E_b}\,\sum_{\sigma=\pm1}\text{PV}\int_0^{k^2}dp\, \frac{Q_\ell(p)}{p-q-\sigma\varpi}
    = \mathcal{L}_\ell(\omega^2,k^2)
    +\frac{2\pi^2\mathcal{A}}{E_b}
    \sum_{\sigma=\pm1}
    Q_\ell(q+\sigma\varpi)
    \log\left|\frac{k^2-q-\sigma\varpi}{q+\sigma\varpi}\right| \, ,
\end{align}
where we have defined
\begin{equation}
\label{eq:Lell_class}
    \mathcal{L}_\ell(\omega^2,k^2)
    \equiv
    \frac{2\pi^2\mathcal{A}}{E_b}
    \sum_{\sigma=\pm1}
    \int_0^{k^2}dp\,
    \frac{Q_\ell(p)-Q_\ell(q+\sigma\varpi)}
    {p-(q+\sigma\varpi)} \, .
\end{equation}
We find that the difference $Q_\ell(p)-Q_\ell(q+\sigma\varpi)$ is exactly divisible by $p-(q+\sigma\varpi)$, so the quotient in \eqref{eq:Lell_class} is a polynomial in $p$ and $(q+\sigma\varpi)$. After performing the $p$ integration and summing over $\sigma=\pm1$, all odd powers of $\varpi$ cancel. Hence $\mathcal{L}_\ell(\omega^2,k^2)$ is a real polynomial of degree $\ell$ in $\omega^2$, and can therefore be cancelled by an appropriate choice of the finite local polynomial $P_\ell(\omega^2,k^2)$ in the classical limit.

The logarithmic term generated by the integral in \eqref{eq:Q_Theta_cl} vanishes in the classical limit because it is $O(\log k / k)$, exactly like the logarithmic term in \eqref{eq:Re_GR_ren_app_3}. Combining everything, in the classical limit we obtain
\begin{equation}
\label{eq:Re_GR_class_match}
    \text{Re}\,\widehat{G}^{R,\text{ren}}_\ell(\omega)
    = P_\ell(\omega^2,k^2)-\mathcal{L}_\ell(\omega^2,k^2) + \cdots \, ,
\end{equation}
where the dots denote terms that vanish in this limit. Matching this expression to the scheme-independent classical result \eqref{eq:Re_chi_int} requires
\begin{equation}
\label{eq:P_class_match}
    P_\ell(\omega^2,k^2) - \mathcal{L}_\ell(\omega^2,k^2) \stackrel{\text{cl}}{\longrightarrow} 0 \, .
\end{equation}
Hence the classical matching condition fixes the classical asymptotics of the finite local counterterms, so that they cancel the polynomial $\mathcal{L}_\ell(\omega^2,k^2)$. The fact that this cancellation is possible is non-trivial, and it is the main result of this appendix: it follows from the fact that, in the classical limit, every contribution to $\text{Re}\,\widehat{G}^{R,\text{ren}}_\ell(\omega)$ in \eqref{eq:Re_GR_ren_app_3} vanishes except for a polynomial term that has exactly the same structure of the finite ambiguity $P_\ell(\omega^2,k^2)$ allowed by the quantum counterterms.

\section{Closed form of the renormalized response}
\label{app:nonpol}
 
Here we evaluate the integral of $R_\ell$ in \eqref{eq:Re_GR_ren},
\begin{equation}
\label{eq:I_ell_sigma}
    \mathcal{I}_{\ell,\sigma}(\omega) \equiv \frac{1}{2\pi^2} \, \text{PV} \int_0^{+\infty} dp \, \frac{R_\ell(p)}{p-a_\sigma} \, , \qquad a_\sigma \equiv q+\sigma\varpi \, ,
\end{equation}
and derive \eqref{eq:GR_closed}. At $\omega=0$ this is the integral \eqref{eq:sI}, $\mathcal{I}_{\ell,\sigma}(0)=\mathcal{I}_\ell$.
 
\paragraph{Reduction to a single integral.}
The function \eqref{eq:Rell} is
\begin{equation}
    R_\ell(p) = \pi^2 \, Q_\ell(p) \, h(p) \, , \qquad h(p) \equiv \coth\!\big(\pi(\sqrt{p}-k)\big) + \coth\!\big(\pi(\sqrt{p}+k)\big) - 2 \, .
\end{equation}
The polynomial \eqref{eq:Qell} has the factor $p-k^2$, which cancels the pole of $h(p)$ at $p=k^2$. We write $Q_\ell(p)=(p-k^2)\,U_\ell(p)$ and divide $U_\ell$ by $p-a_\sigma$,
\begin{equation}
    U_\ell(p) = U_\ell(a_\sigma) + (p-a_\sigma) \, \tilde{U}_\ell(p) \, ,
\end{equation}
where $\tilde{U}_\ell(p)$ is a polynomial in $p$ whose coefficients are polynomials in $a_\sigma$ and $k^2$. Then
\begin{equation}
\label{eq:J_def}
    \mathcal{I}_{\ell,\sigma}(\omega) = \frac{1}{2} \, U_\ell(a_\sigma) \, J(a_\sigma) + \frac{1}{2} \int_0^{+\infty} dp \, \tilde{U}_\ell(p) \, (p-k^2) \, h(p) \, ,
\end{equation}
with
\begin{equation}
    J(a) \equiv \text{PV} \int_0^{+\infty} dp \, \frac{p-k^2}{p-a} \, h(p) \, .
\end{equation}
The second term has no pole. It is a sum of polynomials in $k^2$; for instance, $\int_0^{\infty} dp \, (p-k^2) \, h(p) = k^4+\frac{2}{3}k^2+\frac{1}{30}$. This term is therefore a polynomial in $a_\sigma$ and $k^2$.
 
\paragraph{The integral $J(a)$.}
For $a<0$ there is no pole in the integration range, and the integral can be computed, after a few changes of variable, in terms of the digamma function. The result is
\begin{equation}
\label{eq:J_result}
    J(a) = \frac{1}{3} - 2k^2 + 2 \, (a-k^2) \, \Big[ \log(-a) - \Psi_k(a) \Big] \, ,
\end{equation}
with $\Psi_k$ defined in \eqref{eq:Psi_k}. Without the principal value, both sides of \eqref{eq:J_result} are analytic in $a$ away from the positive real axis, with the logarithm and the square root on their principal branches, so \eqref{eq:J_result} holds for all such $a$. For $a>0$ the principal value is the average of the values above and below the axis, which is the real part. For real $a$ of either sign, then,
\begin{equation}
    J(a) = \frac{1}{3} - 2k^2 + 2 \, (a-k^2) \, \Big[ \log|a| - \text{Re}\,\Psi_k(a) \Big] \, .
\end{equation}
 
\paragraph{Result.}
Since $(a-k^2)\,U_\ell(a)=Q_\ell(a)$, we obtain
\begin{equation}
\label{eq:I_result}
    \mathcal{I}_{\ell,\sigma}(\omega) = Q_\ell(a_\sigma) \, \Big[ \log|a_\sigma| - \text{Re}\,\Psi_k(a_\sigma) \Big] + \mathrm{Pol} \, ,
\end{equation}
where $\mathrm{Pol}$ denotes a polynomial in $a_\sigma$ and $k^2$, which after the sum over $\sigma=\pm1$ is a polynomial in $\omega^2$ and $k^2$ and can be absorbed into the finite local counterterms. In \eqref{eq:Re_GR_ren} the logarithm of \eqref{eq:I_result} cancels that of the first term, and
\begin{equation}
\label{eq:GR_nonP}
    \text{Re}\,\widehat{G}^{R,\text{ren}}_\ell(\omega) = \frac{2\pi^2\mathcal{A}}{E_b} \sum_{\sigma=\pm1} Q_\ell(a_\sigma) \, \Big[ \log\mu - \text{Re}\,\Psi_k(a_\sigma) \Big] + \mathrm{Pol} \, .
\end{equation}
The same steps apply for $\text{Im}\,\omega>0$, where $a_\sigma$ is complex, starting from \eqref{eq:GR_FT_SP}. In \eqref{eq:basic_integral_app} $\log(\epsilon|a|)$ becomes $\log(-\epsilon a)$, the integral of $R_\ell$ is given by \eqref{eq:J_result}, and the counterterms are the same. This gives \eqref{eq:GR_closed}, and its form at real frequency follows with $\varpi\to\varpi+i0^+$. As a check, using $\text{Im}\,\psi(iy)=\frac{\pi}{2}\coth(\pi y)+\frac{1}{2y}$ one finds, for $a_\sigma>0$,
\begin{equation}
    \text{Im} \left[ Q_\ell(a_\sigma) \, \Psi_k(a_\sigma) \right] = -\frac{\sigma}{2\pi} \, F_\ell(a_\sigma) \, ,
\end{equation}
which reproduces \eqref{eq:Im_GR_FT}. The analytic structure of the retarded Green's function in the complex frequency domain is analyzed in detail in \cite{Emparan:2026ta}.
 
\paragraph{Thresholds.}
For $q\neq0$, near an isolated threshold $a_\sigma=0$ we expand \eqref{eq:GR_closed} using
\begin{equation}
    \psi'(ik)+\psi'(-ik)+\frac{1}{k^2} = -\frac{\pi^2}{\sinh^2(\pi k)} \, , \qquad Q_\ell(a_{\sigma}=0) = -\frac{\sinh^2(\pi k)}{\pi^2} \, |\Gamma(\ell+1+ik)|^4 \, ,
\end{equation}
and find
\begin{equation}
\label{eq:threshold_sqrt}
    \widehat{G}^{R,\text{ren}}_\ell(\omega) = -\frac{2\pi^2\mathcal{A}}{E_b} \, |\Gamma(\ell+1+ik)|^4 \, \sqrt{-a_\sigma} \, \big[ 1 + O(a_\sigma) \big] + \left( \text{terms analytic in } a_\sigma \right) .
\end{equation}
There is no logarithm. The square root is real for $a_\sigma<0$ and imaginary for $a_\sigma>0$: it contributes to the conservative response on one side of the threshold and to the dissipative response on the other. This non-analytic behavior cannot be removed by local counterterms, so the conservative response cannot vanish identically as a function of frequency, unlike in the classical case \eqref{eq:Re_chi_int}.
 
\paragraph{Limits.}
For $\ell \ge 1$, when $|\varpi|$ is much larger than $2k$ and $\ell(\ell+1)$, the arguments of both digamma functions in \eqref{eq:Psi_k} are large and $\text{Re}\,\Psi_k(a_\sigma)\simeq\log|\varpi|$.
 
In the static limit $a_\sigma=q$, and for $\ell\geq1$ \eqref{eq:GR_nonP} gives \eqref{eq:lambda_ell_closed}, with $L_\ell=\text{Re}\,\Psi_k(q)$ and $k=\sqrt{2E/E_b}$. For $k\gg\ell(\ell+1)$ we have $\text{Re}\,\psi(i\sqrt{q}+ik)\simeq\log(2k)$ and $\text{Re}\,\psi(i\sqrt{q}-ik)\simeq-\gamma_E$ and hence
\begin{equation}
    L_\ell \simeq \log(2k) - \gamma_E \, , \qquad \lambda_\ell \simeq \text{B}_\ell \, \log\frac{\mu}{2k} + \mathrm{Pol} = -\frac{\text{B}_\ell}{2} \, \log\frac{8E}{E_b \, \mu^2} + \mathrm{Pol} \, ,
\end{equation}
where the term $\gamma_E \, \text{B}_\ell$ is a polynomial in $E$ and has been absorbed in $\mathrm{Pol}$. The logarithm of $E$ arises only at these energies. At lower energies $L_\ell$ tends to a constant, with the cusp in \eqref{eq:threshold_sqrt} at $q=0$,
\begin{equation}
    L_\ell = L_\ell\big|_{q=0} - \frac{\pi^2}{\sinh^2(\pi k)} \, \sqrt{-q} \; \Theta(-q) + O(q) \, ,
\end{equation}
and there is no logarithmic singularity at $E=0$ or at the threshold. At $E=0$, $L_\ell=2\psi\big(\sqrt{\ell(\ell+1)}\big)+1/\sqrt{\ell(\ell+1)}\simeq\log\ell(\ell+1)$. In both regimes the dependence on the energy is not polynomial, so it cannot be removed by local counterterms. These limits are summarized in \eqref{eq:Psi_scales}.

\bibliography{Paper/QLove}

\providecommand{\href}[2]{#2}\begingroup\raggedright\begin{thebibliography}{10}

\bibitem{Binnington:2009bb}
T.~Binnington and E.~Poisson, \emph{{Relativistic theory of tidal Love numbers}}, \href{https://doi.org/10.1103/PhysRevD.80.084018}{\emph{Phys. Rev. D} {\bfseries 80} (2009) 084018} [\href{https://arxiv.org/abs/0906.1366}{{\ttfamily 0906.1366}}].

\bibitem{Damour:2009vw}
T.~Damour and A.~Nagar, \emph{{Relativistic tidal properties of neutron stars}}, \href{https://doi.org/10.1103/PhysRevD.80.084035}{\emph{Phys. Rev. D} {\bfseries 80} (2009) 084035} [\href{https://arxiv.org/abs/0906.0096}{{\ttfamily 0906.0096}}].

\bibitem{Kol:2011vg}
B.~Kol and M.~Smolkin, \emph{{Black hole stereotyping: Induced gravito-static polarization}}, \href{https://doi.org/10.1007/JHEP02(2012)010}{\emph{JHEP} {\bfseries 02} (2012) 010} [\href{https://arxiv.org/abs/1110.3764}{{\ttfamily 1110.3764}}].

\bibitem{Hui:2020xxx}
L.~Hui, A.~Joyce, R.~Penco, L.~Santoni and A.~R. Solomon, \emph{{Static response and Love numbers of Schwarzschild black holes}}, \href{https://doi.org/10.1088/1475-7516/2021/04/052}{\emph{JCAP} {\bfseries 04} (2021) 052} [\href{https://arxiv.org/abs/2010.00593}{{\ttfamily 2010.00593}}].

\bibitem{Rodriguez:2026iot}
M.~J. Rodr{\'\i}guez, L.~Santoni and A.~R. Solomon, \emph{{Love numbers of black holes and compact objects}},  \href{https://arxiv.org/abs/2604.08653}{{\ttfamily 2604.08653}}.

\bibitem{Chakraborty:2026qru}
S.~Chakraborty and P.~Pani, \emph{{Tidal Response of Compact Objects}},  \href{https://arxiv.org/abs/2604.08679}{{\ttfamily 2604.08679}}.

\bibitem{LeTiec:2020bos}
A.~Le~Tiec, M.~Casals and E.~Franzin, \emph{{Tidal Love Numbers of Kerr Black Holes}}, \href{https://doi.org/10.1103/PhysRevD.103.084021}{\emph{Phys. Rev. D} {\bfseries 103} (2021) 084021} [\href{https://arxiv.org/abs/2010.15795}{{\ttfamily 2010.15795}}].

\bibitem{LeTiec:2020spy}
A.~Le~Tiec and M.~Casals, \emph{{Spinning Black Holes Fall in Love}}, \href{https://doi.org/10.1103/PhysRevLett.126.131102}{\emph{Phys. Rev. Lett.} {\bfseries 126} (2021) 131102} [\href{https://arxiv.org/abs/2007.00214}{{\ttfamily 2007.00214}}].

\bibitem{Chia:2020yla}
H.~S. Chia, \emph{{Tidal deformation and dissipation of rotating black holes}}, \href{https://doi.org/10.1103/PhysRevD.104.024013}{\emph{Phys. Rev. D} {\bfseries 104} (2021) 024013} [\href{https://arxiv.org/abs/2010.07300}{{\ttfamily 2010.07300}}].

\bibitem{Ivanov:2022qqt}
M.~M. Ivanov and Z.~Zhou, \emph{{Vanishing of Black Hole Tidal Love Numbers from Scattering Amplitudes}}, \href{https://doi.org/10.1103/PhysRevLett.130.091403}{\emph{Phys. Rev. Lett.} {\bfseries 130} (2023) 091403} [\href{https://arxiv.org/abs/2209.14324}{{\ttfamily 2209.14324}}].

\bibitem{Charalambous:2021mea}
P.~Charalambous, S.~Dubovsky and M.~M. Ivanov, \emph{{On the Vanishing of Love Numbers for Kerr Black Holes}}, \href{https://doi.org/10.1007/JHEP05(2021)038}{\emph{JHEP} {\bfseries 05} (2021) 038} [\href{https://arxiv.org/abs/2102.08917}{{\ttfamily 2102.08917}}].

\bibitem{Charalambous:2022rre}
P.~Charalambous, S.~Dubovsky and M.~M. Ivanov, \emph{{Love symmetry}}, \href{https://doi.org/10.1007/JHEP10(2022)175}{\emph{JHEP} {\bfseries 10} (2022) 175} [\href{https://arxiv.org/abs/2209.02091}{{\ttfamily 2209.02091}}].

\bibitem{Hui:2021vcv}
L.~Hui, A.~Joyce, R.~Penco, L.~Santoni and A.~R. Solomon, \emph{{Ladder symmetries of black holes. Implications for love numbers and no-hair theorems}}, \href{https://doi.org/10.1088/1475-7516/2022/01/032}{\emph{JCAP} {\bfseries 01} (2022) 032} [\href{https://arxiv.org/abs/2105.01069}{{\ttfamily 2105.01069}}].

\bibitem{Kehagias:2024rtz}
A.~Kehagias and A.~Riotto, \emph{{Black holes in a gravitational field: the non-linear static love number of Schwarzschild black holes vanishes}}, \href{https://doi.org/10.1088/1475-7516/2025/05/039}{\emph{JCAP} {\bfseries 05} (2025) 039} [\href{https://arxiv.org/abs/2410.11014}{{\ttfamily 2410.11014}}].

\bibitem{Gounis:2024hcm}
L.~R. Gounis, A.~Kehagias and A.~Riotto, \emph{{The vanishing of the non-linear static love number of Kerr black holes and the role of symmetries}}, \href{https://doi.org/10.1088/1475-7516/2025/03/002}{\emph{JCAP} {\bfseries 03} (2025) 002} [\href{https://arxiv.org/abs/2412.08249}{{\ttfamily 2412.08249}}].

\bibitem{Lupsasca:2025pnt}
A.~Lupsasca, \emph{{Why there is no Love in black holes}},  \href{https://arxiv.org/abs/2506.05298}{{\ttfamily 2506.05298}}.

\bibitem{Parra-Martinez:2025bcu}
J.~Parra-Martinez and A.~Podo, \emph{{Naturalness of vanishing black-hole tides}},  \href{https://arxiv.org/abs/2510.20694}{{\ttfamily 2510.20694}}.

\bibitem{Elkhidir:2026lgp}
A.~Elkhidir, G.~Martin, J.~Parra-Martinez and M.~V.~S. Saketh, \emph{{Quantum Love Numbers are Non-Zero}},  \href{https://arxiv.org/abs/2609.25875}{{\ttfamily 2609.25875}}.

\bibitem{Kim:2020dif}
J.-W. Kim and M.~Shim, \emph{{Quantum corrections to tidal Love number for Schwarzschild black holes}}, \href{https://doi.org/10.1103/PhysRevD.104.046022}{\emph{Phys. Rev. D} {\bfseries 104} (2021) 046022} [\href{https://arxiv.org/abs/2011.03337}{{\ttfamily 2011.03337}}].

\bibitem{Emparan:2017qxd}
R.~Emparan, A.~Fernandez-Pique and R.~Luna, \emph{{Geometric polarization of plasmas and Love numbers of AdS black branes}}, \href{https://doi.org/10.1007/JHEP09(2017)150}{\emph{JHEP} {\bfseries 09} (2017) 150} [\href{https://arxiv.org/abs/1707.02777}{{\ttfamily 1707.02777}}].

\bibitem{Franzin:2024cah}
E.~Franzin, A.~M. Frassino and J.~V. Rocha, \emph{{Tidal Love numbers of static black holes in anti-de Sitter}}, \href{https://doi.org/10.1007/JHEP12(2024)224}{\emph{JHEP} {\bfseries 12} (2025) 224} [\href{https://arxiv.org/abs/2410.23545}{{\ttfamily 2410.23545}}].

\bibitem{Cardoso:2018ptl}
V.~Cardoso, M.~Kimura, A.~Maselli and L.~Senatore, \emph{{Black Holes in an Effective Field Theory Extension of General Relativity}}, \href{https://doi.org/10.1103/PhysRevLett.121.251105}{\emph{Phys. Rev. Lett.} {\bfseries 121} (2018) 251105} [\href{https://arxiv.org/abs/1808.08962}{{\ttfamily 1808.08962}}].

\bibitem{DeLuca:2022tkm}
V.~De~Luca, J.~Khoury and S.~S.~C. Wong, \emph{{Implications of the weak gravity conjecture for tidal Love numbers of black holes}}, \href{https://doi.org/10.1103/PhysRevD.108.044066}{\emph{Phys. Rev. D} {\bfseries 108} (2023) 044066} [\href{https://arxiv.org/abs/2211.14325}{{\ttfamily 2211.14325}}].

\bibitem{Barbosa:2025uau}
S.~Barbosa, P.~Brax, S.~Fichet and L.~de~Souza, \emph{{Running Love numbers and the Effective Field Theory of gravity}}, \href{https://doi.org/10.1088/1475-7516/2025/07/071}{\emph{JCAP} {\bfseries 07} (2025) 071} [\href{https://arxiv.org/abs/2501.18684}{{\ttfamily 2501.18684}}].

\bibitem{Cano:2025zyk}
P.~A. Cano, \emph{{Love numbers beyond GR from the modified Teukolsky equation}}, \href{https://doi.org/10.1007/JHEP07(2025)152}{\emph{JHEP} {\bfseries 07} (2025) 152} [\href{https://arxiv.org/abs/2502.20185}{{\ttfamily 2502.20185}}].

\bibitem{Barbosa:2026qcv}
S.~Barbosa, S.~Fichet and L.~de~Souza, \emph{{Running Love numbers of charged black holes}}, \href{https://doi.org/10.1088/1475-7516/2026/09/009}{\emph{JCAP} {\bfseries 09} (2026) 009} [\href{https://arxiv.org/abs/2602.00349}{{\ttfamily 2602.00349}}].

\bibitem{Noumi:2026shc}
T.~Noumi and S.~S.~C. Wong, \emph{{Extremal Love: tidal/electromagnetic deformability, logarithmic running and the weak gravity conjecture}}, \href{https://doi.org/10.1007/JHEP08(2026)033}{\emph{JHEP} {\bfseries 08} (2026) 033} [\href{https://arxiv.org/abs/2601.20962}{{\ttfamily 2601.20962}}].

\bibitem{Emparan:2026wlm}
R.~Emparan, L.~Lehner and S.~Trezzi, \emph{{Love at the String Scale: Tidal Deformability Across the Black Hole-String Transition}},  \href{https://arxiv.org/abs/2608.23077}{{\ttfamily 2608.23077}}.

\bibitem{Iliesiu:2020qvm}
L.~V. Iliesiu and G.~J. Turiaci, \emph{{The statistical mechanics of near-extremal black holes}}, \href{https://doi.org/10.1007/JHEP05(2021)145}{\emph{JHEP} {\bfseries 05} (2021) 145} [\href{https://arxiv.org/abs/2003.02860}{{\ttfamily 2003.02860}}].

\bibitem{Mertens:2022irh}
T.~G. Mertens and G.~J. Turiaci, \emph{{Solvable models of quantum black holes: a review on Jackiw\textendash{}Teitelboim gravity}}, \href{https://doi.org/10.1007/s41114-023-00046-1}{\emph{Living Rev. Rel.} {\bfseries 26} (2023) 4} [\href{https://arxiv.org/abs/2210.10846}{{\ttfamily 2210.10846}}].

\bibitem{Bai:2023hpd}
Y.~Bai and M.~Korwar, \emph{{Near-extremal charged black holes: greybody factors and evolution}}, \href{https://doi.org/10.1007/JHEP03(2023)151}{\emph{JHEP} {\bfseries 03} (2023) 151} [\href{https://arxiv.org/abs/2301.07739}{{\ttfamily 2301.07739}}].

\bibitem{Brown:2024ajk}
A.~R. Brown, L.~V. Iliesiu, G.~Penington and M.~Usatyuk, \emph{{The evaporation of charged black holes}}, \href{https://doi.org/10.1007/JHEP01(2026)109}{\emph{JHEP} {\bfseries 01} (2026) 109} [\href{https://arxiv.org/abs/2411.03447}{{\ttfamily 2411.03447}}].

\bibitem{Maulik:2025hax}
S.~Maulik, X.~Meng and L.~A. Pando~Zayas, \emph{{Quantum-corrected Hawking radiation from near-extremal Kerr-Newman black holes}}, \href{https://doi.org/10.1007/JHEP02(2026)205}{\emph{JHEP} {\bfseries 02} (2026) 205} [\href{https://arxiv.org/abs/2501.08252}{{\ttfamily 2501.08252}}].

\bibitem{Emparan:2025sao}
R.~Emparan, \emph{{Quantum cross-section of near-extremal black holes}}, \href{https://doi.org/10.1007/JHEP04(2025)122}{\emph{JHEP} {\bfseries 04} (2025) 122} [\href{https://arxiv.org/abs/2501.17470}{{\ttfamily 2501.17470}}].

\bibitem{Biggs:2025nzs}
A.~Biggs, \emph{{Following the state of an evaporating charged black hole into the quantum gravity regime}},  \href{https://arxiv.org/abs/2503.02051}{{\ttfamily 2503.02051}}.

\bibitem{Lin:2025wof}
G.~Lin, L.~V. Iliesiu and M.~Usatyuk, \emph{{The evaporation of black holes in supergravity}}, \href{https://doi.org/10.1007/JHEP08(2025)220}{\emph{JHEP} {\bfseries 08} (2025) 220} [\href{https://arxiv.org/abs/2504.21077}{{\ttfamily 2504.21077}}].

\bibitem{Li:2025vcm}
R.~Li, Z.-X. Man and J.~Wang, \emph{{Decoherence of quantum superpositions in near-extremal Reissner-Nordstr{\"o}m black holes with quantum gravity corrections}}, \href{https://doi.org/10.1007/JHEP08(2025)079}{\emph{JHEP} {\bfseries 08} (2025) 079} [\href{https://arxiv.org/abs/2505.07480}{{\ttfamily 2505.07480}}].

\bibitem{Emparan:2025qqf}
R.~Emparan and S.~Trezzi, \emph{{Quantum transparency of near-extremal black holes}}, \href{https://doi.org/10.1007/JHEP10(2025)023}{\emph{JHEP} {\bfseries 10} (2025) 023} [\href{https://arxiv.org/abs/2507.03398}{{\ttfamily 2507.03398}}].

\bibitem{Bhattacharjee:2025wfv}
A.~Bhattacharjee and M.~Saha, \emph{{Quantum evolution of de Sitter black holes near extremality}}, \href{https://doi.org/10.1007/JHEP04(2026)129}{\emph{JHEP} {\bfseries 04} (2026) 129} [\href{https://arxiv.org/abs/2510.18035}{{\ttfamily 2510.18035}}].

\bibitem{Biggs:2026zlp}
A.~Biggs and S.~Trezzi, \emph{{Not all black holes decohere quantum superpositions}},  \href{https://arxiv.org/abs/2605.23880}{{\ttfamily 2605.23880}}.

\bibitem{Emparan:2026ta}
R.~Emparan, S.~Murthy and S.~Trezzi, \emph{{Quantum Quasinormal Modes of Near-Extremal Black Holes}}, {\emph{to appear} (2026) }.

\bibitem{Balasubramanian:1998de}
V.~Balasubramanian, P.~Kraus, A.~E. Lawrence and S.~P. Trivedi, \emph{{Holographic probes of anti-de Sitter space-times}}, \href{https://doi.org/10.1103/PhysRevD.59.104021}{\emph{Phys. Rev. D} {\bfseries 59} (1999) 104021} [\href{https://arxiv.org/abs/hep-th/9808017}{{\ttfamily hep-th/9808017}}].

\bibitem{Klebanov:1999tb}
I.~R. Klebanov and E.~Witten, \emph{{AdS / CFT correspondence and symmetry breaking}}, \href{https://doi.org/10.1016/S0550-3213(99)00387-9}{\emph{Nucl. Phys. B} {\bfseries 556} (1999) 89} [\href{https://arxiv.org/abs/hep-th/9905104}{{\ttfamily hep-th/9905104}}].

\bibitem{deHaro:2000vlm}
S.~de~Haro, S.~N. Solodukhin and K.~Skenderis, \emph{{Holographic reconstruction of space-time and renormalization in the AdS / CFT correspondence}}, \href{https://doi.org/10.1007/s002200100381}{\emph{Commun. Math. Phys.} {\bfseries 217} (2001) 595} [\href{https://arxiv.org/abs/hep-th/0002230}{{\ttfamily hep-th/0002230}}].

\bibitem{Balasubramanian:1999re}
V.~Balasubramanian and P.~Kraus, \emph{{A Stress tensor for Anti-de Sitter gravity}}, \href{https://doi.org/10.1007/s002200050764}{\emph{Commun. Math. Phys.} {\bfseries 208} (1999) 413} [\href{https://arxiv.org/abs/hep-th/9902121}{{\ttfamily hep-th/9902121}}].

\bibitem{Bhattacharjee:2026nmo}
A.~Bhattacharjee, M.~Saha and M.~V.~S. Saketh, \emph{{Quantum Love and ringdown near extremality}},  \href{https://arxiv.org/abs/2610.03708}{{\ttfamily 2610.03708}}.

\bibitem{Ivanov:2022hlo}
M.~M. Ivanov and Z.~Zhou, \emph{{Revisiting the matching of black hole tidal responses: A systematic study of relativistic and logarithmic corrections}}, \href{https://doi.org/10.1103/PhysRevD.107.084030}{\emph{Phys. Rev. D} {\bfseries 107} (2023) 084030} [\href{https://arxiv.org/abs/2208.08459}{{\ttfamily 2208.08459}}].

\bibitem{Chakraborty:2025wvs}
S.~Chakraborty, V.~De~Luca, L.~Gualtieri and P.~Pani, \emph{{Dynamical Love numbers of black holes: Theory and gravitational waveforms}}, \href{https://doi.org/10.1103/fr3y-s1sz}{\emph{Phys. Rev. D} {\bfseries 112} (2025) 104015} [\href{https://arxiv.org/abs/2507.22994}{{\ttfamily 2507.22994}}].

\bibitem{Son:2002sd}
D.~T. Son and A.~O. Starinets, \emph{{Minkowski space correlators in AdS / CFT correspondence: Recipe and applications}}, \href{https://doi.org/10.1088/1126-6708/2002/09/042}{\emph{JHEP} {\bfseries 09} (2002) 042} [\href{https://arxiv.org/abs/hep-th/0205051}{{\ttfamily hep-th/0205051}}].

\bibitem{Charalambous:2024tdj}
P.~Charalambous, \emph{{Love numbers and Love symmetries for p-form and gravitational perturbations of higher-dimensional spherically symmetric black holes}}, \href{https://doi.org/10.1007/JHEP04(2024)122}{\emph{JHEP} {\bfseries 04} (2024) 122} [\href{https://arxiv.org/abs/2402.07574}{{\ttfamily 2402.07574}}].

\bibitem{Iqbal:2009fd}
N.~Iqbal and H.~Liu, \emph{{Real-time response in AdS/CFT with application to spinors}}, \href{https://doi.org/10.1002/prop.200900057}{\emph{Fortsch. Phys.} {\bfseries 57} (2009) 367} [\href{https://arxiv.org/abs/0903.2596}{{\ttfamily 0903.2596}}].

\bibitem{Skenderis:2002wp}
K.~Skenderis, \emph{{Lecture notes on holographic renormalization}}, \href{https://doi.org/10.1088/0264-9381/19/22/306}{\emph{Class. Quant. Grav.} {\bfseries 19} (2002) 5849} [\href{https://arxiv.org/abs/hep-th/0209067}{{\ttfamily hep-th/0209067}}].

\bibitem{Skenderis:2008dg}
K.~Skenderis and B.~C. van Rees, \emph{{Real-time gauge/gravity duality: Prescription, Renormalization and Examples}}, \href{https://doi.org/10.1088/1126-6708/2009/05/085}{\emph{JHEP} {\bfseries 05} (2009) 085} [\href{https://arxiv.org/abs/0812.2909}{{\ttfamily 0812.2909}}].

\bibitem{Saad:2019lba}
P.~Saad, S.~H. Shenker and D.~Stanford, \emph{{JT gravity as a matrix integral}},  \href{https://arxiv.org/abs/1903.11115}{{\ttfamily 1903.11115}}.

\bibitem{Iliesiu:2019lfc}
L.~V. Iliesiu, \emph{{On 2D gauge theories in Jackiw-Teitelboim gravity}},  \href{https://arxiv.org/abs/1909.05253}{{\ttfamily 1909.05253}}.

\bibitem{Nian:2025oei}
J.~Nian, L.~A. Pando~Zayas and C.-Y. Yue, \emph{{Quantum Corrections in the Low-Temperature Fluid/Gravity Correspondence}},  \href{https://arxiv.org/abs/2510.15411}{{\ttfamily 2510.15411}}.

\bibitem{PandoZayas:2025snm}
L.~A. Pando~Zayas and J.~Zhang, \emph{{One-loop Corrected Holographic Shear Viscosity to Entropy Density Ratio at Low Temperatures}},  \href{https://arxiv.org/abs/2510.16100}{{\ttfamily 2510.16100}}.

\bibitem{Cremonini:2025yqe}
S.~Cremonini, L.~Li, X.-L. Liu and J.~Nian, \emph{{Quantum Corrections to $\eta/s$ from JT Gravity}},  \href{https://arxiv.org/abs/2510.21602}{{\ttfamily 2510.21602}}.

\bibitem{Gouteraux:2025exs}
B.~Gout{\'e}raux, D.~M. Ramirez and C.~Supiot, \emph{{Schwarzian quantum corrections to shear correlators of the near-extremal Reissner-Nordstr{\"o}m-AdS black hole}},  \href{https://arxiv.org/abs/2512.19642}{{\ttfamily 2512.19642}}.

\bibitem{Kanargias:2025vul}
A.~Kanargias, E.~Kiritsis, S.~Murthy, O.~Papadoulaki and A.~P. Porfyriadis, \emph{{Holographic shear correlators at low temperatures, and quantum $\eta/s$}},  \href{https://arxiv.org/abs/2512.20443}{{\ttfamily 2512.20443}}.

\bibitem{Maulik:2026dro}
S.~Maulik, L.~A. Pando~Zayas and J.~Zhang, \emph{{Low-Temperature Holographic Conductivity}},  \href{https://arxiv.org/abs/2607.12064}{{\ttfamily 2607.12064}}.

\bibitem{Das:2026ujn}
S.~Das, S.~Maulik, L.~A. Pando~Zayas and J.~Zhang, \emph{{Low-temperature Quantum-corrected Holographic Transport with Momentum Relaxation}},  \href{https://arxiv.org/abs/2607.12066}{{\ttfamily 2607.12066}}.

\bibitem{Danielson:2022tdw}
D.~L. Danielson, G.~Satishchandran and R.~M. Wald, \emph{{Black holes decohere quantum superpositions}}, \href{https://doi.org/10.1142/S0218271822410036}{\emph{Int. J. Mod. Phys. D} {\bfseries 31} (2022) 2241003} [\href{https://arxiv.org/abs/2205.06279}{{\ttfamily 2205.06279}}].

\bibitem{Danielson:2022sga}
D.~L. Danielson, G.~Satishchandran and R.~M. Wald, \emph{{Killing horizons decohere quantum superpositions}}, \href{https://doi.org/10.1103/PhysRevD.108.025007}{\emph{Phys. Rev. D} {\bfseries 108} (2023) 025007} [\href{https://arxiv.org/abs/2301.00026}{{\ttfamily 2301.00026}}].

\bibitem{Gralla:2023oya}
S.~E. Gralla and H.~Wei, \emph{{Decoherence from horizons: General formulation and rotating black holes}}, \href{https://doi.org/10.1103/PhysRevD.109.065031}{\emph{Phys. Rev. D} {\bfseries 109} (2024) 065031} [\href{https://arxiv.org/abs/2311.11461}{{\ttfamily 2311.11461}}].

\bibitem{Biggs:2024dgp}
A.~Biggs and J.~Maldacena, \emph{{Comparing the decoherence effects due to black holes versus ordinary matter}},  \href{https://arxiv.org/abs/2405.02227}{{\ttfamily 2405.02227}}.

\bibitem{Heydeman:2020hhw}
M.~Heydeman, L.~V. Iliesiu, G.~J. Turiaci and W.~Zhao, \emph{{The statistical mechanics of near-BPS black holes}}, \href{https://doi.org/10.1088/1751-8121/ac3be9}{\emph{J. Phys. A} {\bfseries 55} (2022) 014004} [\href{https://arxiv.org/abs/2011.01953}{{\ttfamily 2011.01953}}].

\bibitem{Pereniguez:2021xcj}
D.~Pere{\~n}iguez and V.~Cardoso, \emph{{Love numbers and magnetic susceptibility of charged black holes}}, \href{https://doi.org/10.1103/PhysRevD.105.044026}{\emph{Phys. Rev. D} {\bfseries 105} (2022) 044026} [\href{https://arxiv.org/abs/2112.08400}{{\ttfamily 2112.08400}}].

\bibitem{Charalambous:2021kcz}
P.~Charalambous, S.~Dubovsky and M.~M. Ivanov, \emph{{Hidden Symmetry of Vanishing Love Numbers}}, \href{https://doi.org/10.1103/PhysRevLett.127.101101}{\emph{Phys. Rev. Lett.} {\bfseries 127} (2021) 101101} [\href{https://arxiv.org/abs/2103.01234}{{\ttfamily 2103.01234}}].

\bibitem{BenAchour:2022uqo}
J.~Ben~Achour, E.~R. Livine, S.~Mukohyama and J.-P. Uzan, \emph{{Hidden symmetry of the static response of black holes: applications to Love numbers}}, \href{https://doi.org/10.1007/JHEP07(2022)112}{\emph{JHEP} {\bfseries 07} (2022) 112} [\href{https://arxiv.org/abs/2202.12828}{{\ttfamily 2202.12828}}].

\bibitem{Berens:2022ebl}
R.~Berens, L.~Hui and Z.~Sun, \emph{{Ladder symmetries of black holes and de Sitter space: love numbers and quasinormal modes}}, \href{https://doi.org/10.1088/1475-7516/2023/06/056}{\emph{JCAP} {\bfseries 06} (2023) 056} [\href{https://arxiv.org/abs/2212.09367}{{\ttfamily 2212.09367}}].

\bibitem{Rai:2024lho}
M.~Rai and L.~Santoni, \emph{{Ladder symmetries and Love numbers of Reissner-Nordstr{\"o}m black holes}}, \href{https://doi.org/10.1007/JHEP07(2024)098}{\emph{JHEP} {\bfseries 07} (2024) 098} [\href{https://arxiv.org/abs/2404.06544}{{\ttfamily 2404.06544}}].

\end{thebibliography}\endgroup

\end{document}